\documentclass[aps,showpacs,amsmath,amssymbd,nofootbib,showkeys,preprint]{revtex4}

\usepackage{graphicx}
\usepackage{color}
\usepackage{natbib}

\def \be  {\begin{equation}}
\def \ee  {\end{equation}}
\def \ee  {\end{equation}}
\def \bea {\begin{eqnarray}}
\def \eea {\end{eqnarray}}

\newcommand{\nn}{\nonumber}

\begin{document}

\preprint{ECTP-2014-05\hspace*{0.5cm}and\hspace*{0.5cm}WLCAPP-2014-05}

\title{Generalized Uncertainty Principle and Recent Cosmic Inflation Observations}

\author{Abdel Nasser TAWFIK\footnote{http://atawfik.net/}}
\affiliation{Egyptian Center for Theoretical Physics (ECTP), Modern University for Technology and Information (MTI), 11571 Cairo, Egypt}
\affiliation{World Laboratory for Cosmology And Particle Physics (WLCAPP), Cairo, Egypt}
\author{Abdel Magied DIAB} 
\affiliation{World Laboratory for Cosmology And Particle Physics (WLCAPP), Cairo, Egypt}

\begin{abstract}

The recent background imaging of cosmic extragalactic polarization (BICEP2) observations are believed as an evidence for the cosmic inflation. BICEP2 provided a first direct evidence for the inflation, determined its energy scale and debriefed witnesses for the quantum gravitational processes. The ratio of scalar-to-tensor fluctuations $r$ which is the canonical measurement of the gravitational waves, was estimated as $r=0.2 _{-0.05}^{+0.07}$. Apparently, this value agrees well with the upper bound value corresponding to PLANCK $r\leq 0.012$ and to WMAP9 experiment $r=0.2$. It is believed that the existence of a minimal length is one of the greatest predictions leading to modifications in the Heisenberg uncertainty principle or a generalization of the uncertainty principle (GUP) at the Planck scale. In the present work, we investigate the possibility of interpreting recent BICEP2 observations through quantum gravity or GUP. We estimate the slow-roll parameters, the tensorial and the scalar density fluctuations which are characterized by the scalar field $\phi$. Taking into account the background (matter and radiation) energy density, $\phi$ is assumed to interact with the gravity and with itself. We first review the Friedmann-Lemaitre-Robertson-Walker (FLRW) Universe and then suggest modification in the Friedmann equation due to GUP. By using a single potential for a chaotic inflation model, various inflationary parameters are estimated and compared with the PLANCK and BICEP2 observations. While GUP is conjectured to break down the expansion of the early Universe (Hubble parameter and scale factor), two inflation potentials based on certain minimal supersymmetric extension of the standard model result in $r$ and spectral index matching well with the observations. Corresponding to BICEP2 observations, our estimation for $r$ depends on the inflation potential and the scalar field. A power-law inflation potential does not.

\end{abstract}

\pacs{98.80.Cq,  04.60.-m, 98.80.Cq}
\keywords{Inflationary universe, Quantum gravity, Early Universe} 

\maketitle
\makeatletter
\let\toc@pre\relax
\let\toc@post\relax
\makeatother 

\section{Introduction} 
\label{intro}

Constrains to the inflationary cosmological models can be set by cosmological observations \cite{Liddle:2003,Linde:2002}. The inflationary expansion not only solves various problems, especially in the early Universe such as the Big Bang cosmology \cite{Starobinsky:1979,Sato:1981,Albrecht:1982,Linde:1982,Guth:1981}, but also provides an explanation for the large-scale structure from the quantum fluctuation of an inflationary field, $\phi$ \cite{Hawking,Guth:1982b,Starobinsky:1982b}. Furthermore, the gravitational waves and the polarization due to the existence of the inflation was discovered in the cosmic microwave background (CMB) \cite{BICEP2}.

Not only the physicists around the world are very aware of the existence of the background imaging of cosmic extragalactic polarization (BICEP2) telescope at the south pole, but the world public as well. It is believed that the BICEP2 observations offer an evidence for the cosmic inflation \cite{BICEP2}. Other confirmations from Planck \cite{Planck,Planck2} and WMAP9 \cite{WMAP} measurements, for instance, are likely in near future. BICEP2 did not only provide the first direct evidence for the inflation, but also determined its energy scale and furthermore debriefed witnesses for the quantum gravitational processes in the inflationary era, in which a primordial density and gravitational wave fluctuations are created from the quantum fluctuations \cite{Mukhanov,Bardeen}. The ratio of scalar-to-tensor fluctuation, $r$, which is  a canonical measurement  of the gravitational waves  \cite{Liddle:2003, Linde:2002}, was estimated by BICEP2, $r=0.2 _{-0.05}^{+0.07}$   \cite{BICEP2}. This value is apparently comparable with the upper bound value corresponding to PLANCK $r\leq 0.012$ and to WMAP9 experiment $r=0.2$. On the other hand, the PLANCK satellite \cite{Planck,Planck2} has reported the scalar spectral index $n_s\approx\,0.96$.

If these observations are true, then the hypothesis that our Universe should go through a period of cosmic inflation will be confirmed and the energy scale of inflation should be very near to the Planck scale \cite{Amaldi}. The large value of tensor-to-scalar ratio, $r$, requires inflation fields as large as the Planck scale. This idea is known as the Lyth bound \cite{Lyth:96,Lyth:98,Green}, which estimates the change of the inflationary field $\Delta \phi$,
\bea
\frac{\Delta \phi}{M_p} &=& \sqrt{\frac{r}{8}} \Delta N,
\eea
where $M_p$ is the Planck mass and $\Delta N$ denotes the number of e-folds corresponding to the observed scales in the CMB left the inflationary horizon. Since the Planckian effects become important and need to be taken into account during the inflation era, as indicated by the Lyth bound, then $\Delta \phi$ should be smaller than or comparable  with the Planck scale $\mid\Delta \phi\mid\ll M_p$. This constrain suggests focusing on concrete inflation field models. In this case, the many corrections suppressed by the Planck scale appear less problematic but come in tension with BICEP2 discovery. Thus, more observations are required to confirm this conclusion.

Various approaches to the quantum gravity (QG) offer quantized description for some problems of gravity, for details readers can consult Ref.  \cite{Tawfik:BH2013}. The effects of minimal length and maximal momentum which are likely applicable at the Planck scale (inflation era) which lead to modifications in the Heisenberg uncertainty principle appear in quadratic and/or linear terms of momentum. These can be implemented at this energy scale. The quadratic GUP was predicted in different theories such as string theory, black hole physics and loop QG \cite{Tawfik:BH2013,Amati88,Amati87,Amati90,Maggiore93,Maggiore94,Kempf,Kempf97,
Kempf2000,Kempf93,Kempf94,Kempf95,Kempf96,Scardigli,Scardigli2009}. The latter, the linear GUP, was introduced by doubly Special Relativity (DSR), which suggests a minimal uncertainty in position and a maximum measurable  momentum \cite{DSR,Smolin,Amelino2002,Tawfik:BH2013}. Accordingly, a minimum measurable length and a maximum measurable momentum \cite{advplb,Das:2010zf,afa2} are simultaneously likely. This offers a major revision of the quantum phenomena \cite{amir,Tawfik:BH2013,Pedram}. This approach has the genetic name, {\it Generalized (gravitational) Uncertainty Principle (GUP)}. Recently, various implications of GUP approaches on different physical systems have been carried out \cite{Tawfik:2013uza,Ali:2013ma,Ali:2013ii,Tawfik:2012he,Tawfik:2012hz,Elmashad:2012mq,DiabBH2013}, App. \ref{GUPy}. 

In the present work, we estimate various inflationary parameters which are characterized by the scalar field $\phi$ and apparently contribute to the total energy density. Taking into account the background (matter and radiation) energy density, the scalar field is assumed to interact with the gravity and with itself. The coupling of $\phi$ to gravity is assumed to result in total inflation energy. We first review the Friedmann-Lemaitre-Robertson-Walker (FLRW) Universe and then suggest modifications in the Friedmann equation due to GUP. Using modified Friedmann equation and a single potential for a chaotic inflation model, the inflationary parameters are estimated and compared with PLANCK and BICEP2 observations.

The applicability of the GUP approaches in estimating inflationary parameters comparable with the recent BICEP2 observations will be discussed.  In section \ref{FRW}, we present Friedmann-Lemaitre-Robertson-Walker (FLRW) Universe and introduce the modification of Friedmann equation due to GUP at planckian scale in matter and radiation background. In section \ref{cos_inflat}, the modified Friedmann equation in cosmic inflation will be introduced. Some inflation potentials for chaotic inflation models will be surveyed. We suggest to implement the single inflationary field $\phi$. In the cosmic inflation models and quantum fluctuations, the inflationary parameters are given in section \ref{Fluctuation}. The discussion and final conclusions will be outlined in section \ref{summary}. Appendix \ref{GUPH} gives details about the higher order GUP with minimum length uncertainty and maximum measurable momentum in  Hilbert space. The applicability of GUP to the cosmic inflation will be elaborated in Appendix  \ref{app:appl}. The modified dispersion relation (MDR) as an alternative to the GUP will be introduced in App. \ref{app:mdr}.

\section{Generalized uncertainty principle in FLRW background} 
\label{FRW}

In $(n+1)$-dimensional FLRW Universe, the metric can be described by the line element as \cite{dinverno}
\bea
ds^2 =\, c^2 d t^2 + a(t)^2 \left(\frac{dr^2}{1-\kappa~r^2} + r^2 \, d \theta ^2 +\, r^2 sin^2 \, d \phi ^2\right), \label{metric}
\eea
where $a(t)$ is the scale factor and $\kappa$ is the curvature constant that measures the spatial flatness $\pm 1$ and $0$. In Einstein-Hilbert space, the action reads 
\bea
S &=& \int \left( \frac{1}{8\, \pi \, G} L_{G} + L_{\phi} \right) d\,\Omega, \label{action}
\eea
where  $d\Omega =d\theta \,+\,sin \theta \,d\phi$ , $G$ is the gravitational constant, $L_G$ is the geometrical Lagrangian related to the line element of the FLRW Universe and $L_{\phi}$ \cite{Rong:2005} is Lagrangian coupled to the scalar field $\phi$  
\bea
L_{\phi}\,= -\left[g^{\mu \, \nu}\, \partial _{\mu} \phi \, \partial_{\nu} \phi  +V(\phi)\right], 
\eea
where $V(\phi)$ is the potential and $g^{\mu\, \nu}$ is diagonal matrix $\,diag\left\lbrace 1, -1, -1, -1\right\rbrace$. Under the assumption of homogeneity and isotropy, a standard simplification of the variables leads to the FLRW  metric, where the gradient of the scalar field vanishes. The integration of the action, Eq. (\ref{action}), over a unit volume results in $L_{\phi}=-\frac{1}{2}a^3\dot{\phi}^2-a^3V(\phi)$. 
Since the FLRW Lagrangian of scalar field evaluated at vanishing mass, results in,
\bea
L &=& \frac{1}{2} a^3\, \dot{\phi}^2 -\frac{3}{8\, \pi\, G}\left(a\, \dot{a}^2 - a\, \kappa \right),
\label{Lagra2}
\eea
and the energy-momentum tensor reads
\bea
T^{\mu \nu} &=& \partial ^{\nu} \frac{\partial L}{\partial (\partial_{\mu}\, \phi)} - g^{\mu\,  \nu} L,
\eea
while the four momentum tensor is given by $P^{\mu}= T^{\mu \, 0}$ and the Hamiltonian constraint $\mathcal{H}=P^0 =T^{0 \, 0}$  
\bea 
\mathcal{H} &=& \pi\, \frac{\partial L}{\partial (\partial _{0} \phi)} - L, \label{Hconst0}
\eea
where the $\pi=\partial L/\partial \dot{\phi}$ is known as the canonical momentum conjugate for the scalar field $\phi$. Thus, the total Hamiltonian is given as 
\bea 
\mathfrak{h}  &=& \int d^3\, x \, \mathcal{H}.
\eea
The scalar field becomes equivalent to a perfect fluid with respectively energy density and pressure 
\bea
\rho &=& \frac{\dot{\phi}}{2}\,+\, V(\phi), \label{rhophi} \\ 
p &=& \frac{\dot{\phi}}{2}\,-\, V(\phi).
\eea

When taking into account the cosmological constant $\Lambda$, then the energy density $\rho \rightarrow \rho+\rho _v$, with $\rho_v=\Lambda/8\, \pi \, G$.
Using Eq. (\ref{Lagra2}) and taking into account Eq. (\ref{rhophi}), the dynamics of such models are summarized in the Hamiltonian constraint 
\bea 
\mathcal{H}&=&-\frac{2\pi \, G}{3}\, \frac{p_{a}^2}{a} -\frac{3}{8\, \pi \, G}\, \kappa a\,+\, a^3 \rho \equiv 0.
\label{Hconst}
\eea
This equation is equivalent to the estimation for FLRW Universe \cite{HAWKING:1986,Roman,Farag2014}, where the momenta $p_a$ associated with the scalar factor are defined as 
\bea
p_{a}&:=& \frac{\partial \mathcal{L}}{\partial \dot{a}} = \frac{-3}{4\, \pi \, G}\, a\, \dot{a}.
\eea
The standard Friedmann equations can be extracted from the equations of motion  which can be derived from the extended Hamiltonian by exchanging the negative sign in Eq. (\ref{Hconst}) in order to estimate the exact form of  Friedmann equations 
\bea
\mathcal{H}_E &=& \frac{2\pi \, G}{3}\, \frac{p_{a}^2}{a} + \frac{3}{8\, \pi \, G}\, \kappa a - a^3 \rho.
\eea
Based on the relationship between the commutation relation and the Poisson bracket which was first proposed by Dirac \cite{Dirac},  we get for two quantum counterparts $\hat{A}$ and $\hat{B}$ and two observables  $A$ and $B$ that
\bea 
[\hat{A},\hat{B}] &=&i\ \hbar \, \lbrace A,B \rbrace.
\eea
In the standard case, the canonical uncertainty relation for variables of scale factor $a$ and momenta $p_a$ satisfies the Poisson bracket $\lbrace a, \,p_a  \, \rbrace=\, 1$. Then, the equations of motion read
\bea 
\dot{a}&=&\,\lbrace a,\,\mathcal{H}_E \rbrace = \lbrace a,\, p_a \rbrace \frac{\partial\mathcal{H}_E}{\partial p_a} = \left( \frac{4\, \pi \, G}{3}\right) \frac{p_a}{a}, \label{adot}\\ 
\dot{p_a}&=&\,\lbrace p_a,\,\mathcal{H}_E \rbrace \,=-\, \lbrace a,\, p_a \rbrace \frac{\partial\mathcal{H}_E}{\partial a} = \left( \frac{2\, \pi \, G}{3}\right) \frac{p_{a}^2}{a^2} -\frac{3}{8\, \pi \, G}\, \kappa \, +\, 3\,a^2 \rho \,+\, a^3 \frac{\partial \rho}{\partial a}. \label{pdot}
\eea
From Eqs. (\ref{adot}) and (\ref{pdot}) and the Hamiltonian constrain, Eq. (\ref{Hconst}), then the Friedmann equation is given as  
\bea 
H^2 &=& 
\left( \frac{8\, \pi \, G}{3}\right) \rho \, -\, \frac{\kappa}{a^2},
\label{fried}
\eea
where $H=\dot{a}/a$ is the Hubble parameter. For a cosmic fluid, the energy density is combined from a contribution due to the inflation $\rho (\phi)$, Eq. (\ref{rhophi}) and another part related to the inclusion of the cosmological constant, $\rho_v$.  

Now, we consider the higher-order GUP in deformed Poisson algebra in order to study classical approaches, such as Friedmann equations, Appendix \ref{GUPy}. We introduce GUP in terms of {\it first order} $\alpha$ \cite{Tawfik:BH2013}. Accordingly, the Poisson bracket between the scale factor $a$ and momenta $p_a$  reads
\bea 
\lbrace a\, ,\, p_a \rbrace = 1 - 2\, \alpha\, p_a.
\eea
We follow the same procedure as in Eq. (\ref{fried}), but for a modified term of QG, we will use the extended Hamiltonian with the Poisson brackets to get the modified equations of motion
\bea
\dot{a}&=& 
\lbrace a,\, p_a \rbrace \frac{\partial\mathcal{H}_E}{\partial p_a}\,=\,(1-2\alpha p_a)\, \frac{4 \pi G}{3} \frac{p_a}{a},  \label{adota}\\
\dot{p_a}&=& 
\lbrace a,\, p_a \rbrace \frac{\partial\mathcal{H}_E}{\partial a}=(1-2\alpha p_a)\,\left(\frac{2 \pi G}{3} \frac{p_a^2}{a^2} -\frac{3}{8 \pi G} \kappa + 3 a^2 \rho +a^3 \frac{d\rho}{da}\right).  \label{pdotp}
\eea
By using Eqs. (\ref{adota}) and (\ref{pdotp}) with the scalar constraint, Eq. (\ref{Hconst}), we obtain the modified Friedmann equation 
\bea
H^2 &=& 
\left(\frac{8 \pi G}{3} \rho - \frac{\kappa}{a^2}\right) \left[1\,- \, \frac{3\,\alpha \, a^2 }{ \pi G}\left(\frac{8 \pi G}{3} \rho - \frac{\kappa}{a^2}\right)^{1/2}\right].
\label{HddDo}
\eea
By considering the standard case, Eq. (\ref{fried}), in which $\alpha$ vanishes for $\kappa=0$, we find that the modified Friedmann  equation reads
\bea
H^2 &=& \frac{8\, \pi \,G}{3} \rho\, \left[1-\, 3\,\alpha\, a^2  \sqrt{\frac{8}{3\, \pi \, G}}\, \rho^{1/2}\right]. \label{FR3}
\eea

\subsection{Bounds on GUP parameter}

The GUP parameter is given as $\alpha=\alpha _0/(M_{p} c)=\alpha _0 \ell _p/\hbar$, where $c,\; \hbar$ and $M_p$ are  speed of light and Planck constant and mass, respectively. The Planck length $\ell _p\, \approx\, 10^{-35}~$m and the Planck energy $M_p c^2 \,\approx \, 10^{19}~$ GeV.  $\alpha _0$, the proportionality constant, is conjectured to be dimensionless \cite{advplb}. In natural units $c=\hbar=1$, $\alpha$ will be in GeV$^{-1}$, while in the physical units, $\alpha$ should be in GeV$^{-1}$ times $c$. The bounds on $\alpha_0$, which was summarized in Ref. \cite{afa2,AFALI2011,DasV2008}, should be a subject of precise astronomical observations, for instance gamma ray bursts \cite{Tawfik:2012hz}. 
\begin{itemize}
\item Other alternatives were provided by the tunnelling current in scanning tunnelling microscope and the potential barrier problem \cite{AFALI2012}, where the energy of the electron beam is close to the Fermi level. We found that the varying tunnelling current relative to its initial value is shifted due to the GUP effect \cite{AFALI2011,AFALI2012}, $\delta I/I_0\approx 2.7 \times 10^{-35}$ times $\alpha _{0} ^{2}\,$. 
In case of electric current density $J$ relative to the wave function $\Psi$, the current accuracy of precision measurements reaches the level of $10^{-5}$. Thus, the upper bound $\alpha_0<10^{17}$. Apparently, $\alpha$ tends to order $10^{-2}~$GeV$^{-1}$ in natural units or $10^{-2}~$GeV$^{-1}$ times $c$ in physical units. This quantum-mechanically-derived bound is consistent with the one at the electroweak scale \cite{AFALI2011,AFALI2012,DasV2008}. Therefore, this could signal an intermediate length scale between the electroweak and the Planck scales  \cite{AFALI2011,AFALI2012,DasV2008}.
\item On the other hand, for a particle with mass $m$ mass, electric charge $e$ affected by a constant magnetic field ${\vec B}=B {\hat z}\approx10~$Tesla, vector potential ${\vec A}= B\,x \,{\hat y}$ and cyclotron frequency $\omega_c = eB/m$, the Landau energy is shifted due to the GUP effect \cite{AFALI2011,AFALI2012} by
\bea
\frac{\Delta E_{n(GUP)}}{E_n} &=& -\sqrt{8\, m}\; \alpha\;  (\hbar\, \omega _c )^{\frac{1}{2}} \,
 \left(n +\frac{1}{2}\right)^{\frac{1}{2}}  \approx  - 10^{-27}\; \alpha_0.
\eea
Thus, we conclude that if $\alpha_0\sim 1$, then $\Delta E_{n(GUP)}/E_n$ is too tiny to be measured. But with the current measurement accuracy of $1$ in $10^3$, the upper bound on $\alpha_0<10^{24}$ leads to $\alpha=10^{-5}$ in natural units or $\alpha=10^{-5}$ times $c$ in the physical units.

\item Similarly, for the Hydrogen atom with Hamiltonian $H=H_0+H_1$, where standard Hamiltonian $H_0=p_0^2/(2m) - k/r$ and the first perturbation Hamiltonian $H_1 = -\alpha\, p_0^3/m$, it can be shown that the GUP effect on the Lamb Shift \cite{AFALI2011,AFALI2012} reads
\bea
\frac{\Delta E_{n(GUP)}}{\Delta E_n} &\approx & 10^{-24}~\alpha_0.
\eea
Again, if $\alpha_0 \sim 1$,  then $\Delta E_{n(GUP)}/E_n$ is too small to be measured, while the current measurement accuracy gives $10^{12}$. Thus, we assume that  $\alpha_0>10^{-10}$.
\end{itemize}

In light of this discussion, should we assume that the dimensionless $\alpha_0$ has the order of unity in natural units, then $\alpha$ equals to the Planck length $\approx\, 10^{-35}~$m. The current experiments seem not be able to register discreteness smaller than about $10^{-3}$-th fm, $\approx\, 10^{-18}~$m \cite{AFALI2011,AFALI2012}. We conclude that the assumption that $\alpha_0\sim 1$ seems to contradict various observations  \cite{Tawfik:2012hz} and  experiments \cite{AFALI2011,AFALI2012}. Therefore, such an assumption should be relaxed to meet the accuracy of the given experiments. Accordingly, the lower bounds on $\alpha$ ranges from $10^{-10}$ to $10^{-2}~$GeV$^{-1}$. This means that $\alpha_0$ ranges between $10^9\, c$ to $10^{17}\, c$.

\subsection{Standard model solution of Universe expansion}

In a toy model \cite{Tawfik:2011gh,Tawfik:2010ht}, the prefect cosmic fluid contributing to the stress tensor $T_{\mu \nu}$ can be characterized by symmetries of the metric, homogeneity and isotropy of the cosmic Universe. Thus, the total stress-energy tensor $T_{\mu \nu}$ must be diagonal and the spatial components will be given as
\bea
T_{\mu \nu}=\textit{diag}(\rho,-p,-p,-p).
\eea
Assuming that all types of energies in the early Universe are heat $Q$ captured in a closed sphere with radius equal to scale factor $a$ of volume $V=4\pi\,a^3/3$, the energy density during the expansion $\rho=U/V$, where $U$ is internal energy \cite{Tawfik:2011gh,Tawfik:2010ht}. The first law of thermodynamic satisfies of the total energy conservation
\bea
dQ &=& d U + p\, dV = 0.
\label{Henergy}
\eea 
By substituting the totally differential of the energy density, $d\,\rho =dU/V-U\,dV/V^2$ into Eq. (\ref{Henergy}), we get 
\bea
d\,\rho &=& -3 \frac{da}{a}(\rho +p).
\eea   
Dividing both sides over $dt$ results in
\bea
\dot{\rho} &=& - 3 H(\rho +p).
\eea
For a very simple equation of state, $\omega=p/\rho$, where $\omega$ is independent of time,  the energy density reads 
\bea
\rho \sim a^{-3(1+\omega)}.
\eea
The radiation-dominated phase is characteristic by $\omega=1/3$ or $p=\rho/3$. Therefore, $\rho \sim a^{-4}$, the scaling factor, $a \sim \textit{const.}\, t^{1/2}$ and the Hubble parameter, $H=1/(2t)$. In the matter-dominated phase, $\omega=0$, i.e. $p \ll \rho$. Therefore, $\rho \sim  a^{-3}$, $a \sim \textit{const.}\, t^{2/3}$ and $H=2/(3t)$.

\begin{figure}[htb] 
\includegraphics[width=5cm,angle=-90]{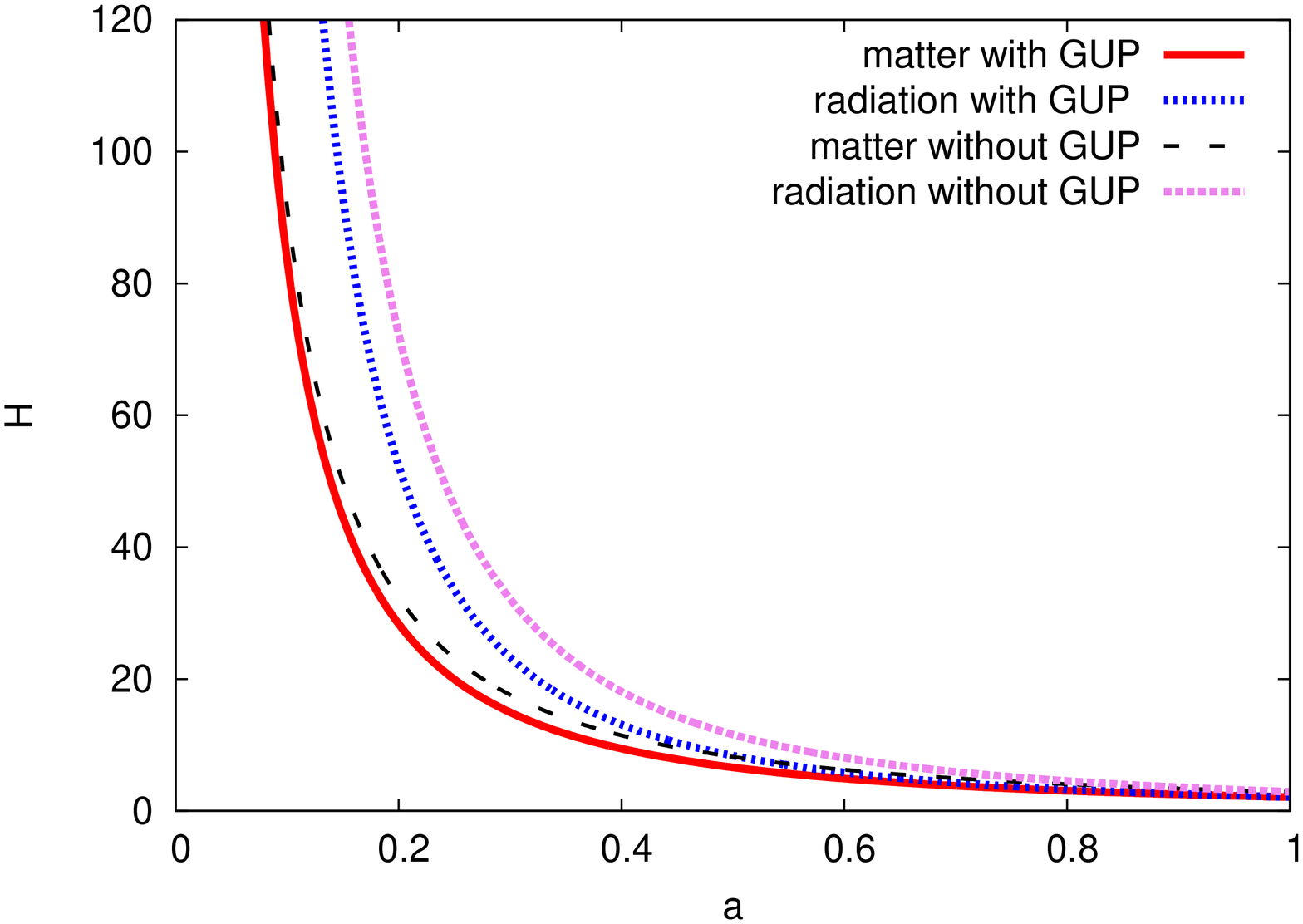}
\includegraphics[width=5cm,angle=-90]{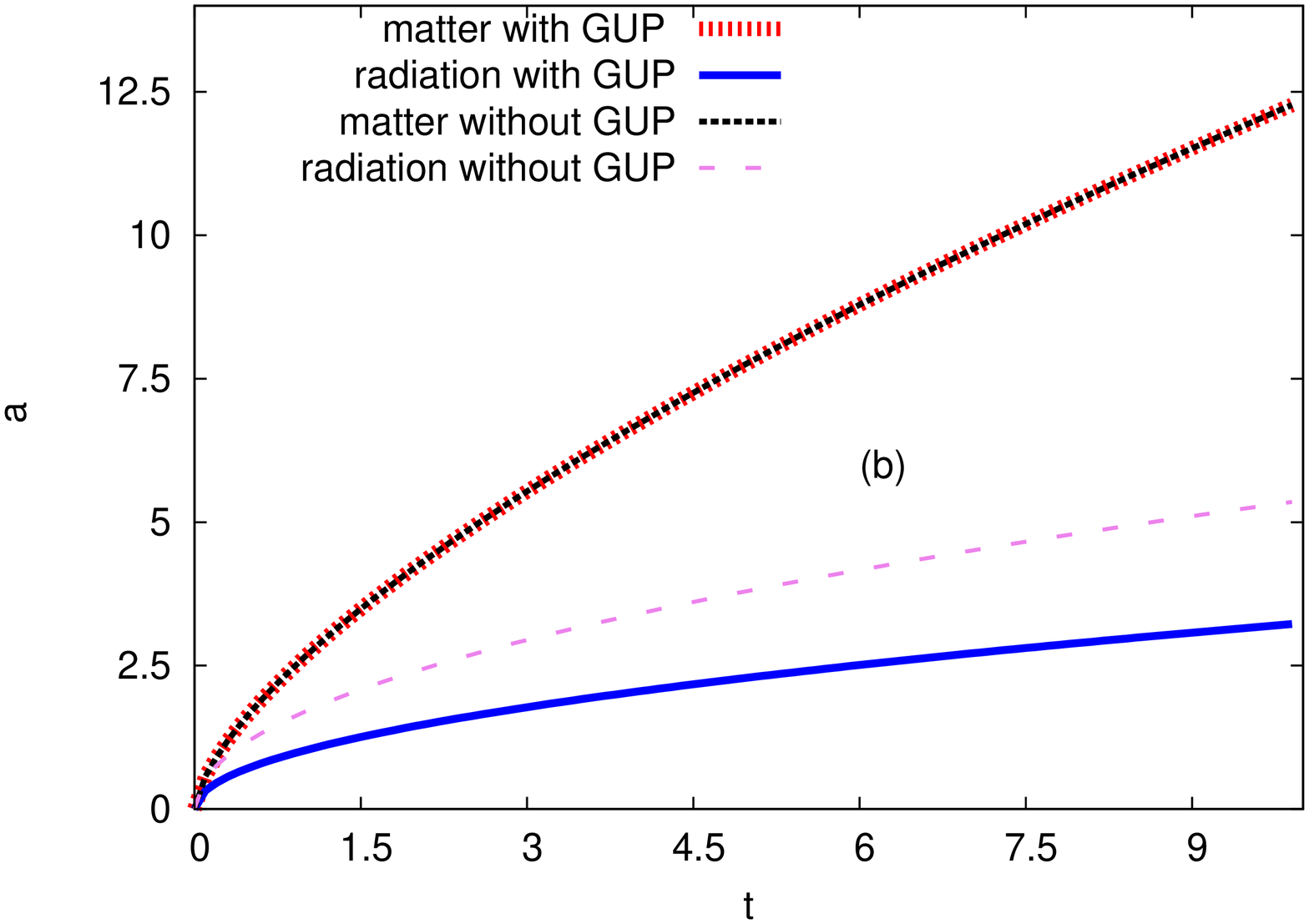}
\caption{(Color online) Left-hand panel (a) presents the variation of the Hubble parameter $H$ with respect to the scale factor $a$. Matter- and radiation-dominated phases with and without GUP are compared with each other. Right-hand panel (b) presents the scale factor $a$ as function of the cosmic time $t$. Various parameters are fixed, $G=1$ and $\alpha =10^{-2}~$GeV$^{-1}$. 
\label{Hubble&scale} 
}
\end{figure}

The left-hand panel (a) of Fig. \ref{Hubble&scale} shows  the Hubble parameter, $H$ in dependence on the scale factor, $a$. The standard (without GUP), Eq. (\ref{fried}) are compared with the modified (with GUP) characterizations of the cosmic fluid, Eq. (\ref{FR3}) in the flat universe. It is obvious that $H$ in both cases  (with/without GUP) diverges at vanishing $a$. This would mean that a singularity exists at the beginning. The GUP has the effect to slightly slow down the expansion rate of the Universe. This is valid for both cases of cosmic background, radiation and matter. 

In the right-panel (b) of Fig. \ref{Hubble&scale}, the dependence of the scale factor, $a$, on the cosmic time, $t$, is given for both cases of cosmic matters, radiation and matter with and without GUP. Apparently, the GUP is not sensitive to the matter-dominated phase but has a clear effect on the radiation-dominated phase. The GUP breaks down the expansion.

\section{Cosmic inflation} 
\label{cos_inflat}

Here, we estimate various inflation parameters,  which are characterized by the scalar field $\phi$ and apparently contribute to the total energy density \cite{Liddle:2003, Linde:2002}. Also, taken into account the background (matter and radiation) energy density, the scalar field is assumed to interact with the gravity and with itself  \cite{Linde:1982,Liddle:1993,Liddle:1995,Liddle:2003,Linde:2002}. In order to reproduce the basics of the field theory, the coupling of $\phi$ to  gravitation results in total inflation energy 
\bea
\frac{1}{2} \left( \dot{\phi}^2 + (\nabla \phi)^2\, \right) + V(\phi).
\label{infl_ener}
\eea
The dynamics of the inflation can be described by  two types of equations:
\begin{itemize}
\item the Friedmann equation, which describes the contraction and expansion of the Universe and
\item the Klein-Gordon equation, which is the simplest equation of motion for a spatially homogeneous scalar field
\bea
\ddot{\phi} + 3 \,H\, \dot{\phi} + \partial _\phi V(\phi) = 0, \label{KG}
\eea
where $\partial_\phi \equiv \partial/\partial \phi$. 
\end{itemize} 
In a flat Universe, $\kappa=0$, the total inflation energy, Eq. (\ref{infl_ener}) and the energy density due to the cosmological constant $\rho_v=\Lambda/8\, \pi \, G$, can be substituted  in the modified Friedmann equation, Eq. (\ref{FR3}), 
\bea
H^2 &=& \frac{8 \pi G}{3} \left[ \frac{\dot{\phi}^2+(\nabla \phi)^2}{2} + V(\phi) +\rho_v \right] \left[1- 3 \alpha  a^2  \sqrt{\frac{8}{3 \pi G}}\left(\frac{\dot{\phi}^2 +(\nabla \phi)^2}{2} + V(\phi) + \rho_v \right)^{1/2}\right]. \hspace*{10mm}
\label{MFR5}
\eea
In rapidly expanding Universe and if the inflation field starts out sufficiently homogeneous, the inflation field becomes minimum, very slow \cite{Liddle:2003, Linde:2002}. This would be modelled by a sphere in a viscous medium, where both the energy densities due to matter $\rho _m$ and radiation $\rho_r$ are neglected 
\bea
(\nabla \phi)^2 &\ll & V(\phi),  \label{eq:inql1} \\
\ddot{\phi} & \ll & 3\, H\, \dot{\phi}, \label{eq:inql2} \\
\dot{\phi}^2 & \ll &  V(\phi). \label{eq:inql3} 
\eea
The first inequality, Eq. (\ref{eq:inql1}), is obtained under the assumption of homogeneity and isotropy of the FLRW Universe \cite{Liddle:2003, Linde:2002}, while the second inequality, Eq. (\ref{eq:inql2}), states that the scalar field changes very slowly so that the acceleration would be neglected \cite{Liddle:2003, Linde:2002}. The third inequality, Eq. (\ref{eq:inql3}), gives a principle condition for the expansion. Accordingly, the kinetic energy is much less than the potential energy \cite{Liddle:2003, Linde:2002}. Apparently, the Universe expansion accelerates \cite{Linde:1982}. Therefore, the modified Friedmann equation, Eq. (\ref{MFR5}) and the Klein-Gordon equation, Eq. (\ref{KG}), respectively read
\bea
H^2 &=& \frac{8 \pi \,G}{3} \left( \, V(\phi) \,+\, \rho_v \right) \left[1-\, 3\,\alpha \, a^2  \sqrt{\frac{8}{3 \pi \, G}}\,\left( \, V(\phi) \,+\, \rho_v \right) ^{1/2} \right], \label{MFE}\\
 \dot{\phi} &=& - \frac{1}{3\, H}\; \partial_\phi V(\phi). 
\eea
The cosmological constant characterizes the minimum mass that is related to the Planck mass $M_P =\sqrt{\hbar c/ G}$. The Planck length $\ell_p = \sqrt{\hbar G/c^{3}}$ \cite{T. Harko:2005} is also related to the mass quanta, where quantized mass \cite{Wesson:2004} is proportional to the GUP parameter $\alpha=\alpha _0/(M_p c)$. The cosmological constant $\Lambda$ is one of the foundation of gravity \cite{Wesson:2004}. It related the Planck (quantum scale) and the Einstein (in cosmological scale) masses, $M_P$ and $M_E$, respectively,  with each other \cite{Wesson:2004}
\bea
M_p &=& \left( \frac{h}{c}\right) \left(\frac{\Lambda}{3}\right)^{1/2},\\
M_E &=& \left( \frac{c^2}{G}\right) \left(\frac{3}{\Lambda}\right)^{1/2}.
\eea

By using natural units $\hbar=c=1$, the modified Friedmann equation, Eq. (\ref{MFE}), becomes 
\bea
H^2\,=\,\frac{4\pi}{3\, \, M_{p}^2}\left\lbrace \left[ V(\phi)\,+\, \frac{3\, M_{p}^4}{4\,\pi}\right] \,-\,3\alpha \, a^2\, \sqrt{\frac{16\,M_{p}^2}{3\,\pi}} \left[ V(\phi)\,+\, \frac{3\, M_{p}^4}{4\,\pi} \right]^{3/2}\right\rbrace.
\label{modified HH}
\eea
There are various inflation models such as chaotic inflation models, which suggest different inflation potentials \cite{Linde:1982,Liddle:1993,Liddle:1995}. Now, it is believed that they are better motivated than other models \cite{Linde:1982,Liddle:1993,Liddle:1995}. In this context, there are two main types of models; one with a single inflation field and the other one combines two inflation fields. Here, we summarize some models requiring a single inflation-field $\phi$ which in some regions satisfies the slow-roll conditions,
\bea
\text{Polynomial chaotic inflation} & & V(\phi)=\frac{1}{2}\,m^2 \phi ^2,  \qquad   V(\phi)=\lambda \phi ^4, \\ 
\text{Power-law inflation} & & V(\phi)=V_0 \exp \left[\sqrt{\frac{16\, \pi\, G}{p}} \phi\right], \\ 
\text{Natural inflation} & & V(\phi)=V_0 \left(1+\cos \frac{\phi}{f}\right), \qquad V(\phi)\propto \phi^{-\beta}.
\eea

Based on this concept, we select three different inflation potential models, Eqs. (\ref{eq:mssm}), (\ref{sdual}) and (\ref{poweri}). The first one is based on certain minimal supersymmetric extensions of the standard model for elementary particles~\cite{allahverdi-2006} and the related effects have been studied, recently~\cite{allahverdi-2006,sanchez-2007}. It has two free parameters, $m$ and $\lambda$, 
\bea
V(\phi) = \left(\frac{m^2}{2}\right)\,\phi^2   - \left(\frac{\sqrt{2\,\lambda\,(n-1)}\,m}{n}\right)\, \phi^n  + \left(\frac{\lambda}{4}\right)\,\phi^{2(n-1)},
\eea
where $n>2$ is an integer. At $n=3$, 
 \bea
V_1(\phi) = \left(\frac{m^2}{2}\right)\,\phi^2   - \left(\frac{2\sqrt{\lambda}\,m}{3}\right)\, \phi^3  + \left(\frac{\lambda}{4}\right)\,\phi^{4}, \label{eq:mssm}
\eea
which is an $\mathcal{S}$-dual inflationary potential \cite{sdual} with a free parameter $f$. The $\mathcal{S}$ duality has its origin in the Dirac quantization condition of the electric and magnetic charges \cite{Montonen:1977}. This would suggest an equivalence in the description of the quantum electrodynamics \cite{Montonen:1977},
\bea
V_2(\phi) &=& V_0\, \text{sech} \left(\frac{\phi}{f}\right).
\label{sdual}
\eea
For a power-law inflation with the free parameter $d$  \cite{Starobinsky,Liddle:1993}, 
\bea
V_3(\phi)=\frac{3 M_{p}^2 d^2}{32 \pi} \left[1-\exp \left(-\frac{16\pi}{3 M_{p}^2 }^{1/2} \phi \right) \right]^2. \label{poweri}
\eea

For these inflation potentials, Eqs. (\ref{eq:mssm}), (\ref{sdual}) and (\ref{poweri}), the inflation parameters such as potential slow-roll parameters $\epsilon$, $\eta$, tensorial $p_t$ and scalar $p_s$ density fluctuations, the ratio of tensor-to-scalar fluctuations $r$, scalar spectral index $n_s$ and the  number of e-folds with the inflation era $\mathcal{N}_e$ can be estimated.

\begin{figure}[htb]
\includegraphics[width=6.5cm,angle=-90]{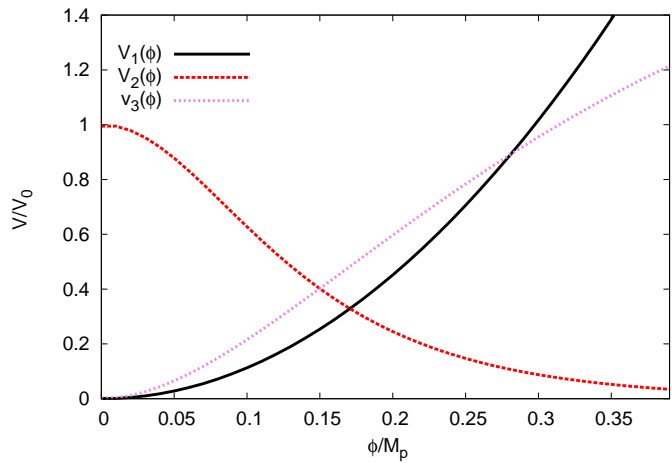}
\caption{(Color online) The variation of  inflation potentials $V/V_0$ is given in dependence on scalar field $\phi/M_p$ at limited free constants. The solid, long-dashed and dotted  line stands for $V_1(\phi)$, $V_2(\phi)$ and $V_3(\phi)$, respectively.
\label{Potentials} 
}
\end{figure}

Fig. \ref{Potentials} shows the variation of the different inflation potentials, Eqs.  (\ref{eq:mssm}), (\ref{sdual}) and (\ref{poweri}), normalized with respect to initial potential $V_0$ with the single inflation field $\phi$ according to Lyth bound during the inflation era \cite{Lyth:96,Lyth:98,Green} and normalized with respect to $M_p$. The inflation field, $\phi \equiv \Delta \phi =(\phi _0 -\phi _{end})$ should be smaller than or comparable with the Planck scale $M_p$. This was confirmed by the BICEP2 observation conditionally with this bound of small scalar field \cite{Lyth:96,Lyth:98,Green}. The potentials, Eqs. (\ref{eq:mssm}) and (\ref{poweri}) increase with $\phi/M_p$, while the third potential, Eq. (\ref{sdual}) , decreases. This means that the latter is finite at vanishing  inflation field, $\phi$, while the earlier vanishes.

\section{Fluctuations and slow-roll parameters in the inflation era} 
\label{Fluctuation}

\begin{figure}[htb]
\includegraphics[width=3.5cm,angle=-90]{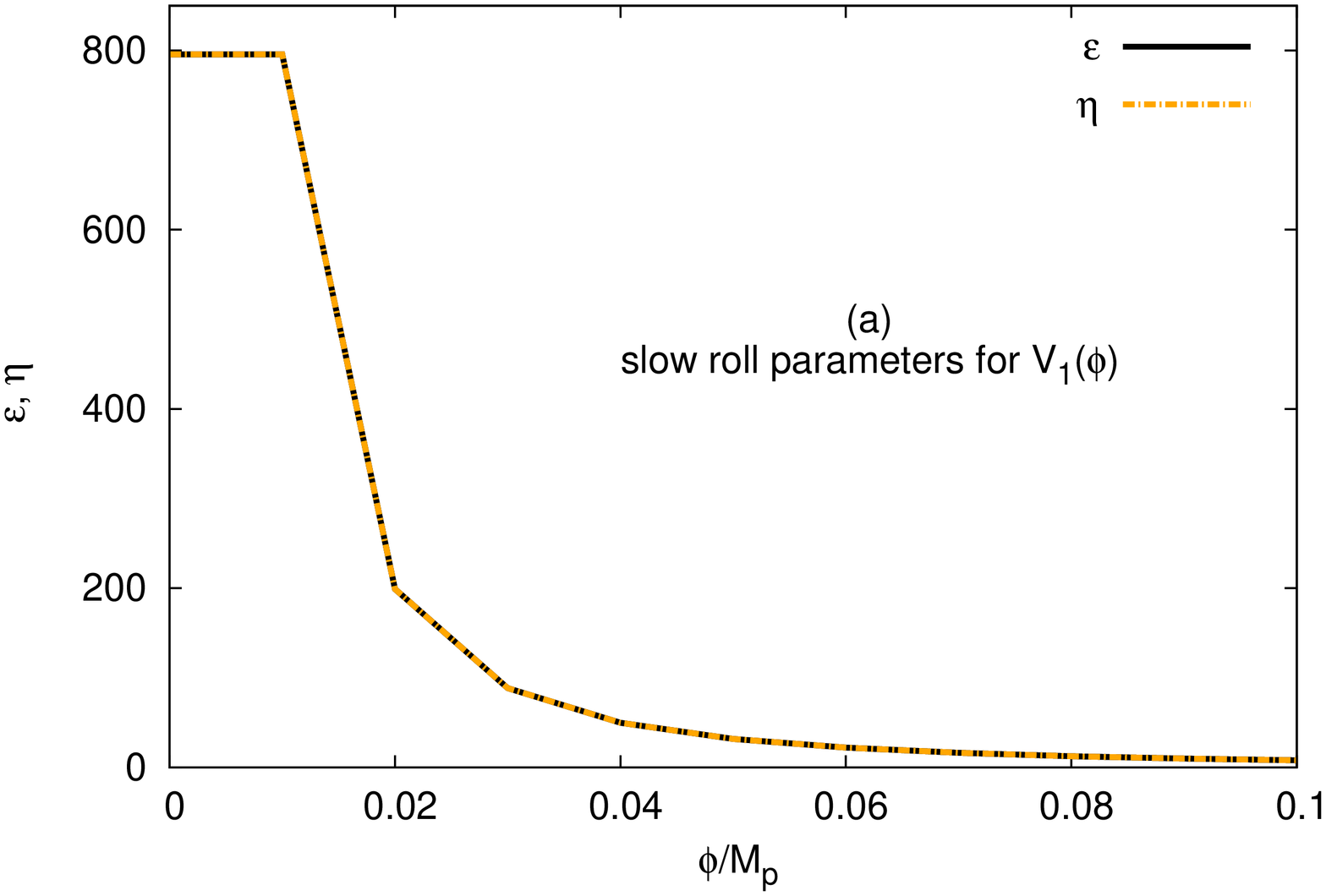}
\includegraphics[width=3.5cm,angle=-90]{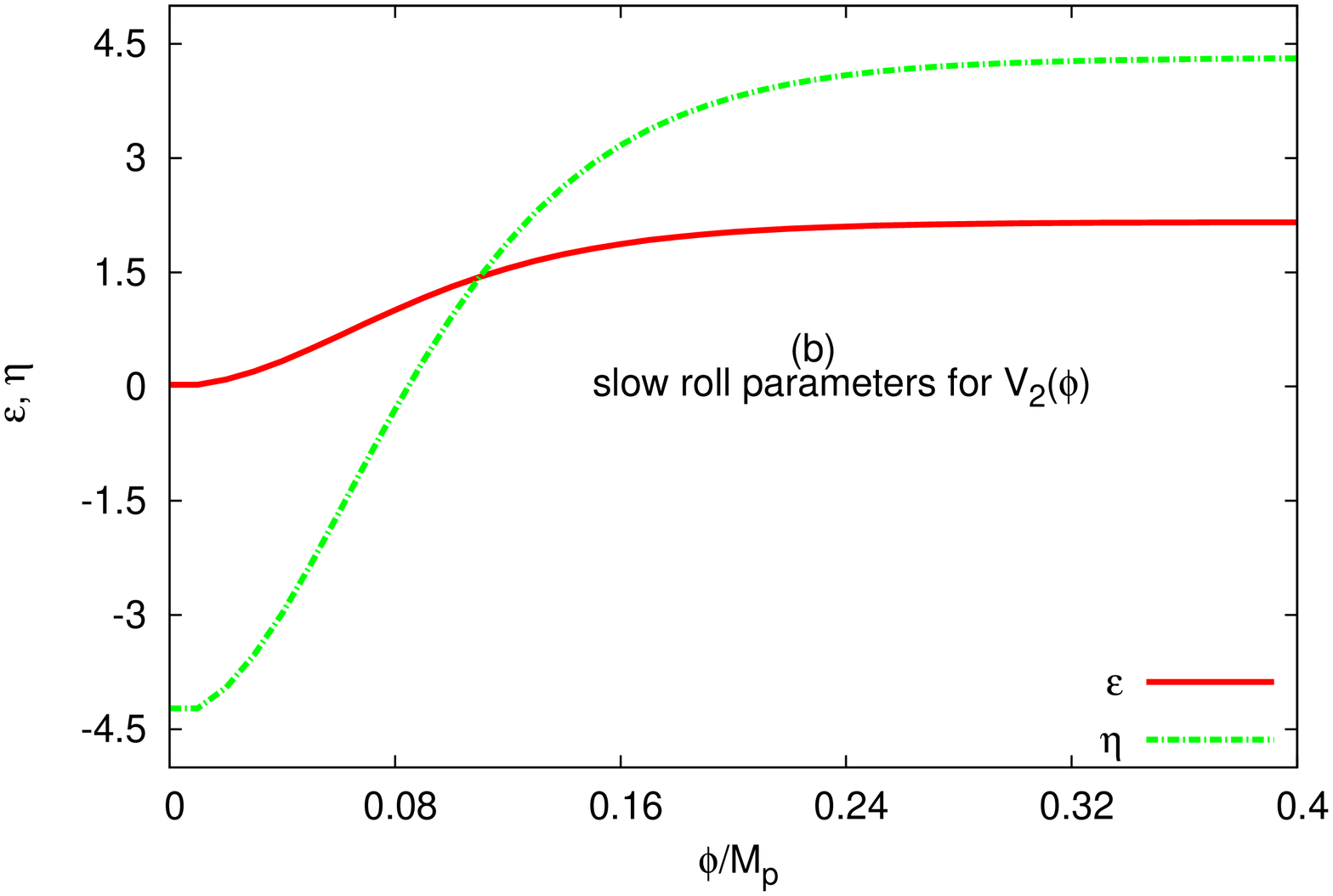}
\includegraphics[width=3.5cm,angle=-90]{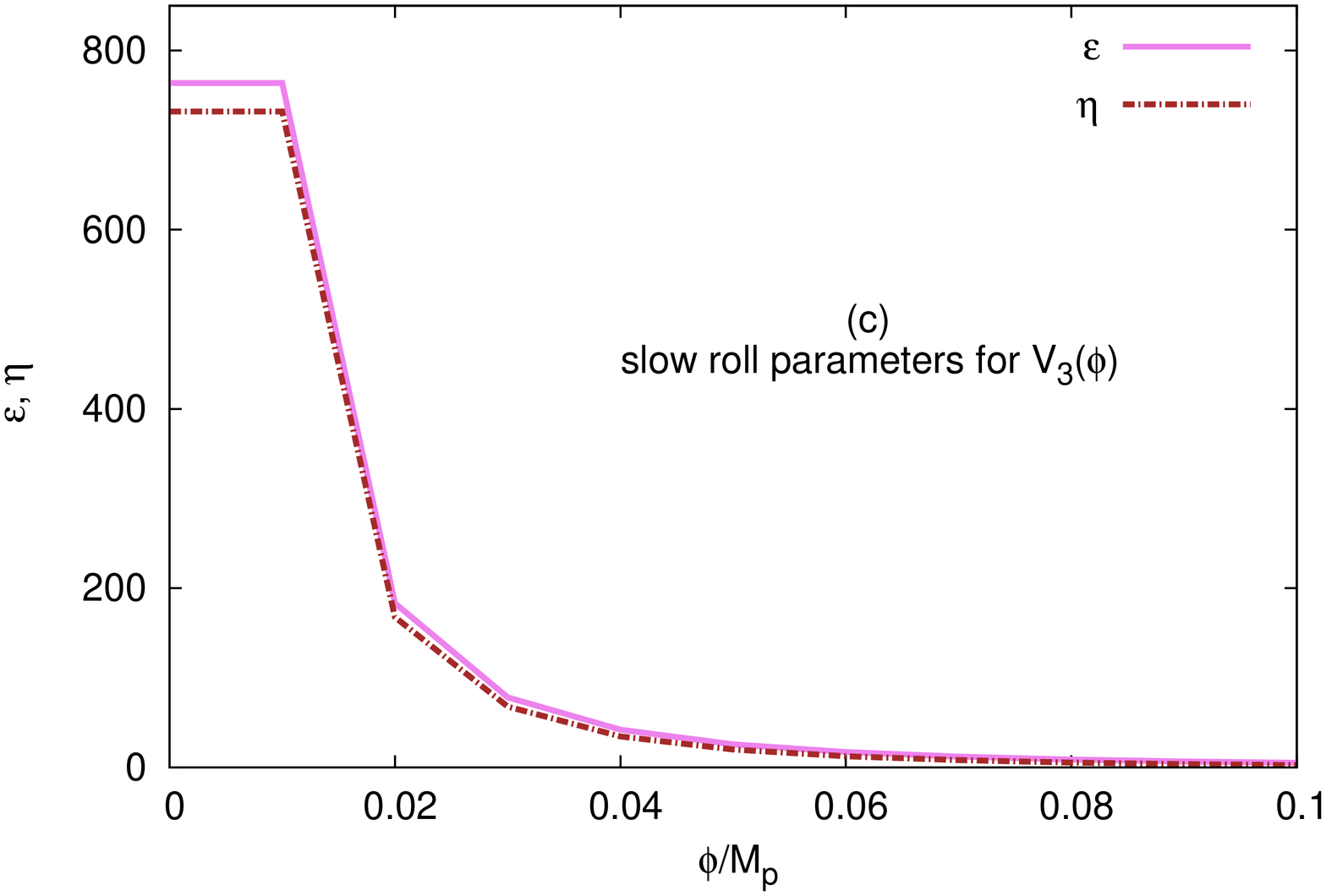}
\caption{(Color online) From the left- (a) middle (b) and right-hand (c) panels present the slow-roll parameters associated with $V_1(\phi)$ from Eq. (\ref{eq:mssm}), $V_2(\phi)$ from Eq. (\ref{sdual}) and $V_3(\phi)$ from (\ref{poweri}) , respectively. The solid and dot-dashed curves represent $\epsilon$ and $\eta$ parameters, respectively. 
\label{slowroll} 
}
\end{figure}

In very early Universe, the scaler field $\phi$ is assumed to derive the inflation  \cite{Linde:1982,Liddle:1993,Liddle:1995}. The main potential slow-roll parameters are given as
\bea
\epsilon &\equiv & \frac{M_{p}^2}{16 \, \pi} \left(\frac{\partial _{\phi} V(\phi)}{ V(\phi)}\right)^2,
\label{paramters1} \\
\eta & \equiv & \frac{M_{p}^2}{8 \pi} \left(\frac{\partial_{\phi}^2 V(\phi)}{ V(\phi)}\right).
\label{paramters2}
\eea
Fig. \ref{slowroll} shows the variation of the potential slow-roll parameters as functions of the scalar field. Various inflation potentials, Eqs. (\ref{eq:mssm}), (\ref{sdual}) and (\ref{poweri}) are used to deduce the slow-roll parameters, Eqs. (\ref{paramters1}) and (\ref{paramters2}). The scalar fields in left- (a) and right-hand panel (c) result in slow-roll parameters, which start from large values at small field. Then, they rapidly decline (vanish) as the scalar field increases. The field presented in the middle panel gives slow-roll parameters with relatively very small values, but seem to remain stable with the field.

\begin{figure}[htb]
\includegraphics[width=3.5cm,angle=-90]{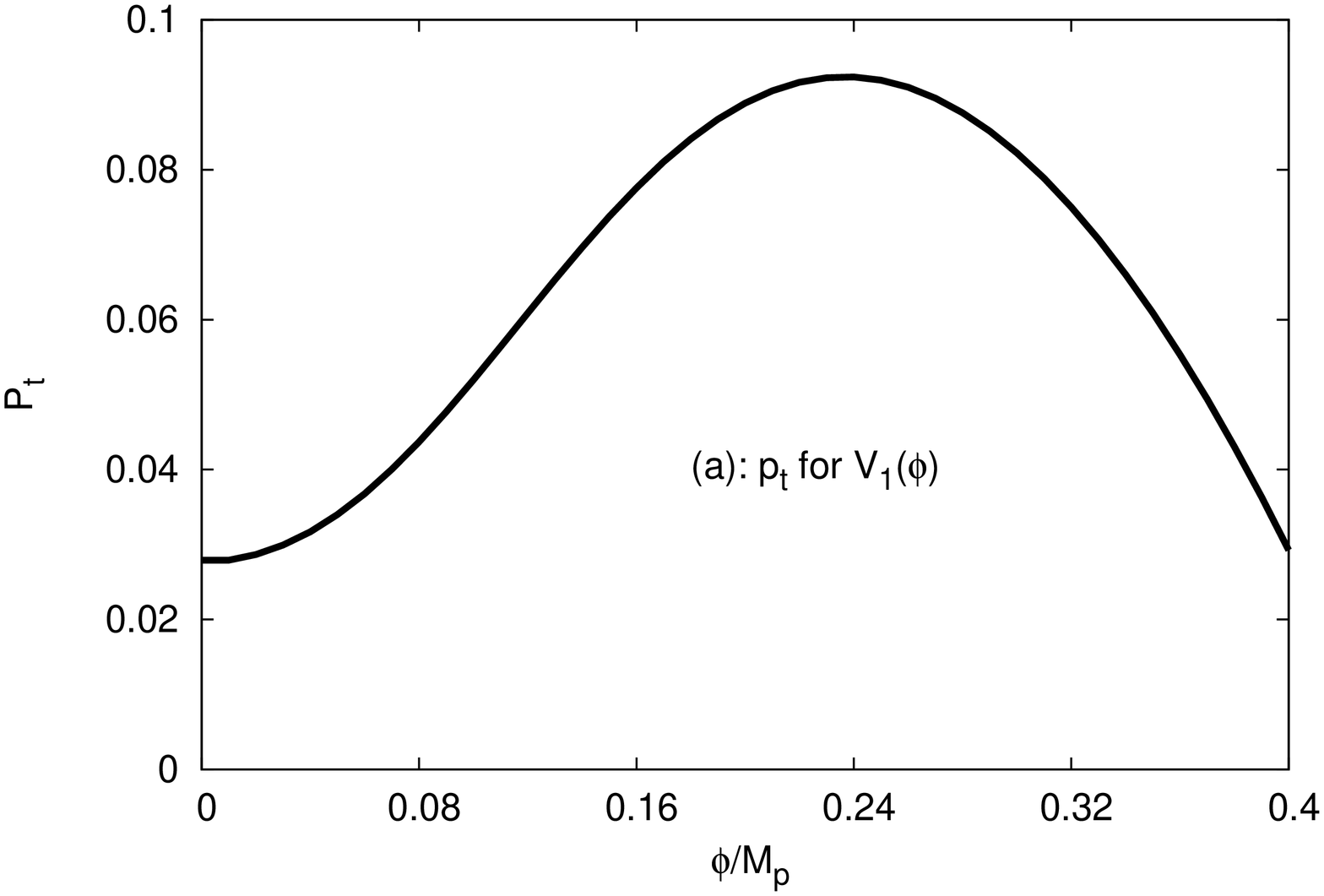}
\includegraphics[width=3.5cm,angle=-90]{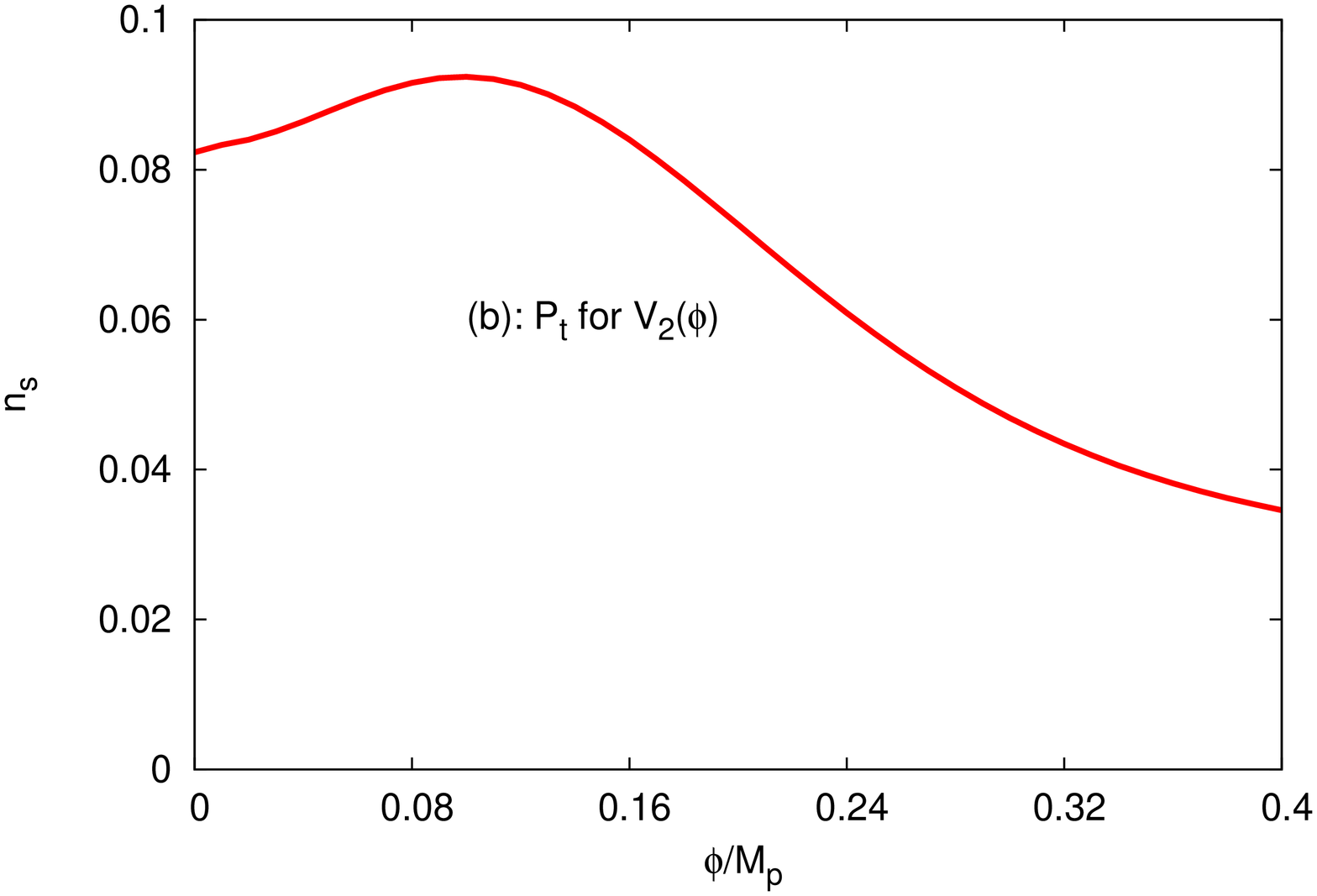}
\includegraphics[width=3.5cm,angle=-90]{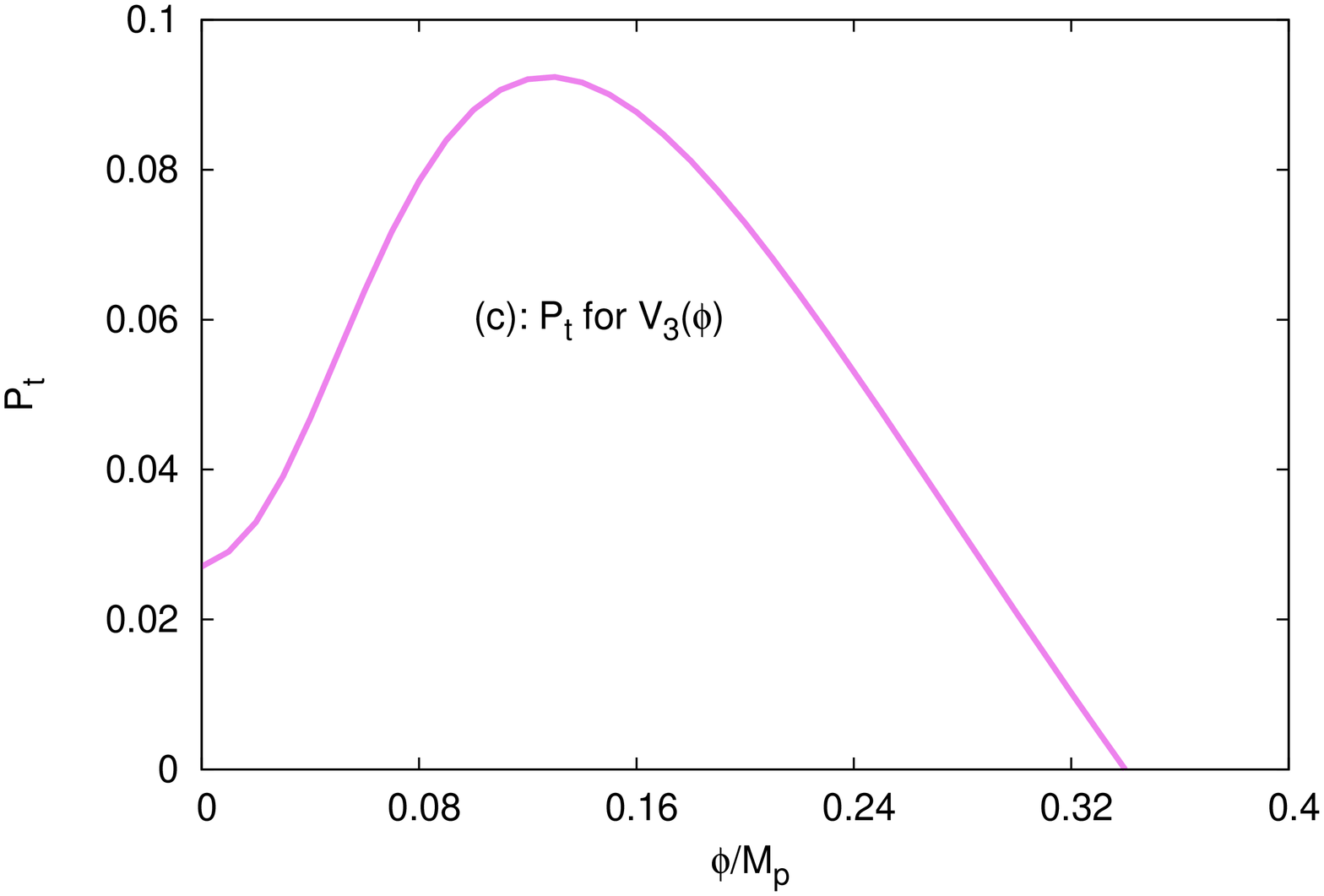}\\
\includegraphics[width=3.5cm,angle=-90]{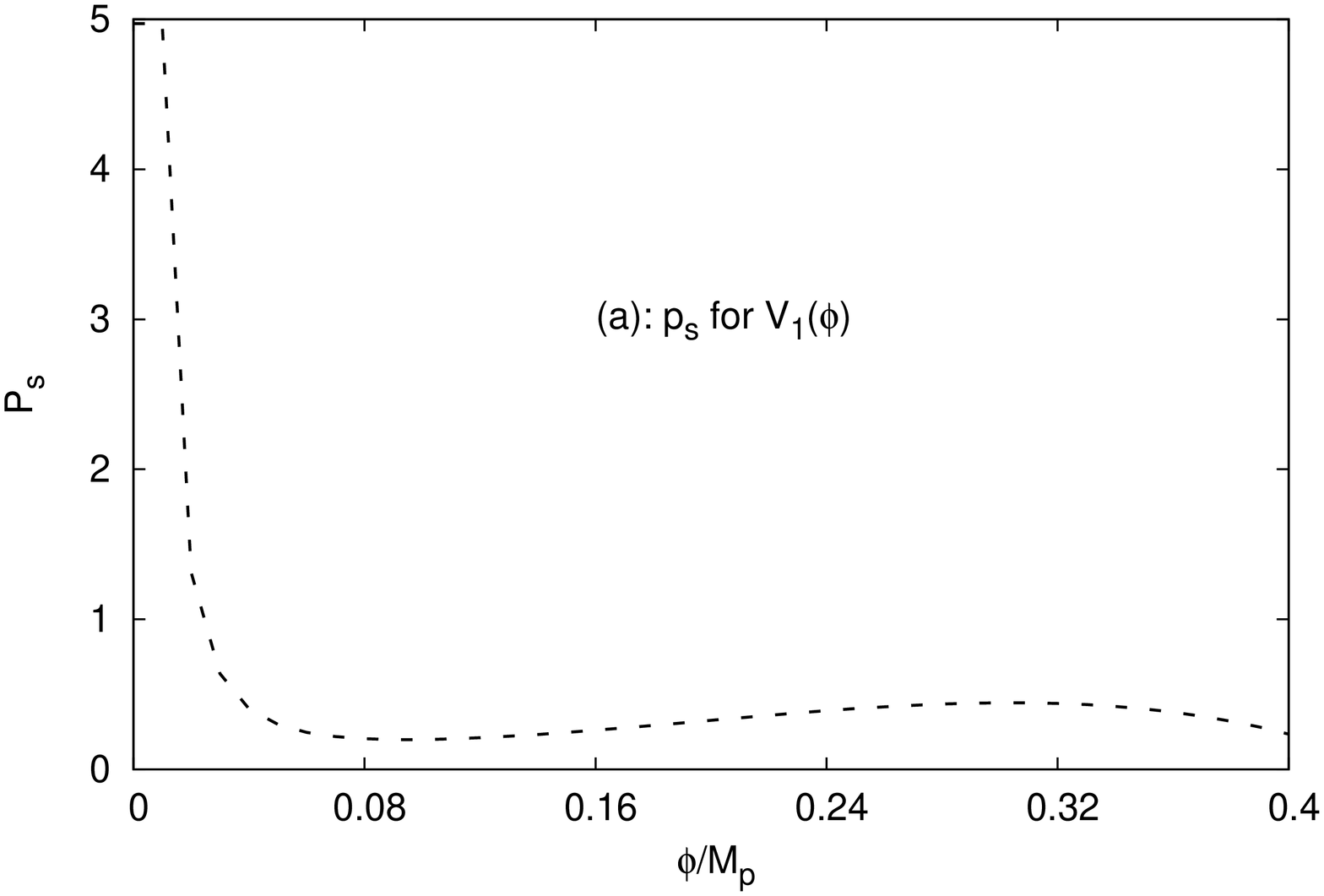}
\includegraphics[width=3.5cm,angle=-90]{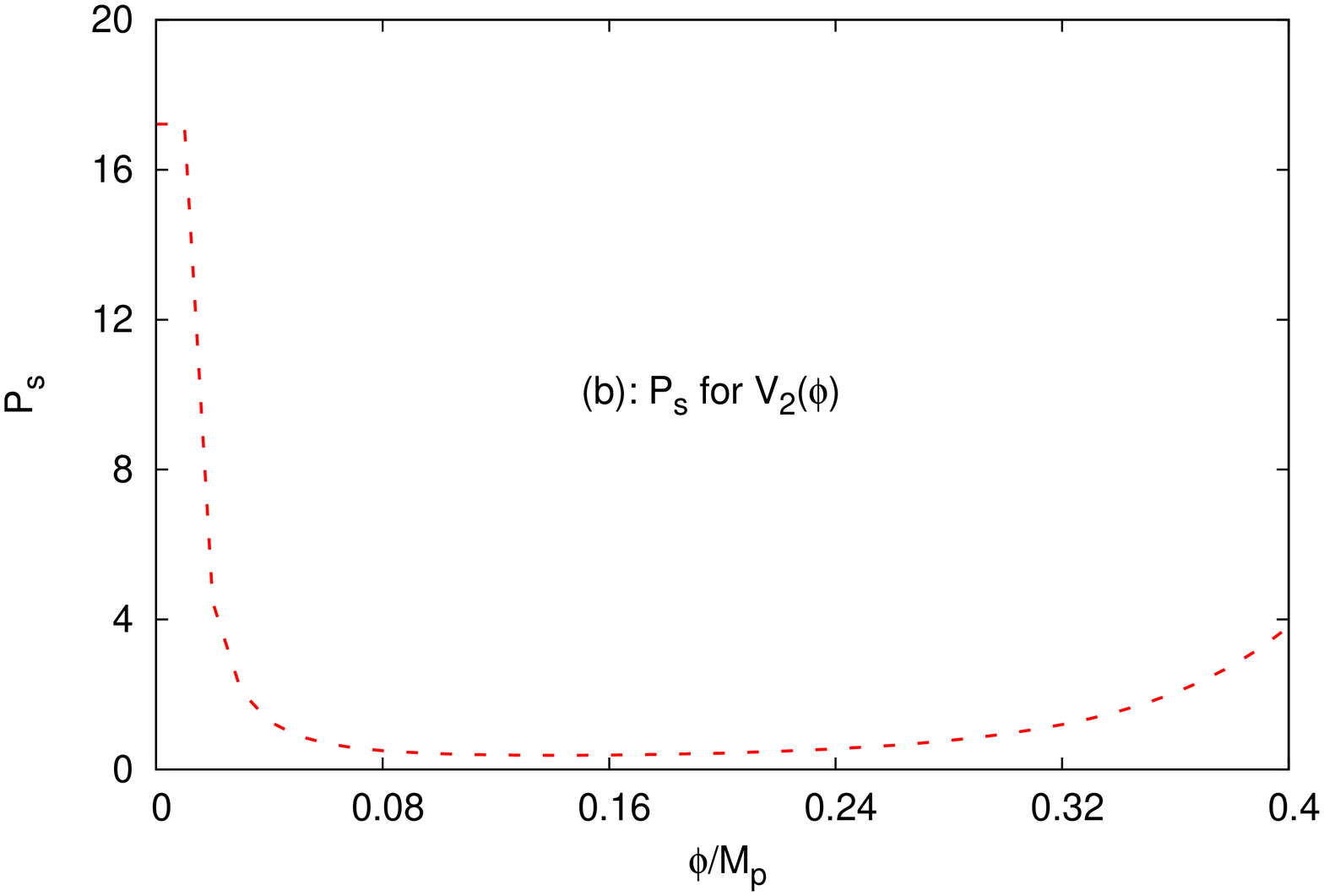}
\includegraphics[width=3.5cm,angle=-90]{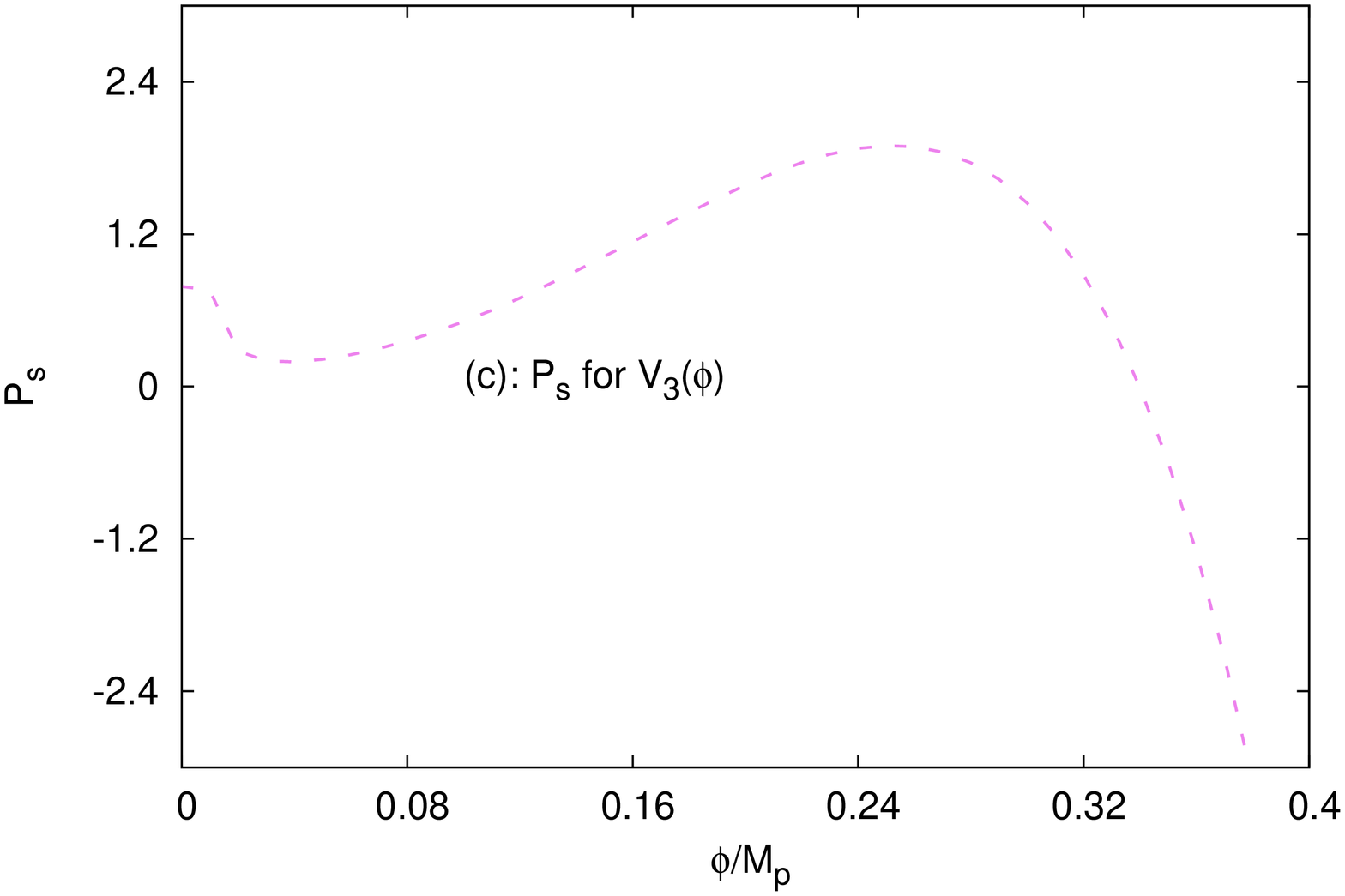}
\caption{(Color online) The top panels show the tonsorial density fluctuations, $p_t$, in dependence on the scalar field $\phi/M_p$. The bottom panels give the scalar density fluctuations $p_s$ in dependence on the scalar field $\phi/M_p$. Each column is associated with an inflation potential model. The results corresponding to $\alpha=10^{-2}~$GeV$^{-1}$ are depicted, only. Exactly same curves are also obtained at $\alpha=10^{-19}~$GeV$^{-1}$ (not shown here).
\label{pt_ps} 
}
\end{figure}

The tensorial and scalar density fluctuations are given as \cite{Linde:1982,Liddle:1993,Liddle:1995}
\begin{eqnarray}
p_t &=& \left(\frac{H}{2\pi}\right)^{2} \left[1-\frac{H}{\Lambda}\,\sin\left(\frac{2\Lambda}{H}\right)\right]=\left(\frac{H}{2\, \pi}\right)^{2} \left[1-\frac{H}{3\, M_p^2}\,\sin\left(\frac{6\, M_p^2}{H}\right)\right],\label{pt} \\
p_s &=& \left(\frac{H}{\dot{\phi}}\right)^2\left(\frac{H}{2\pi}\right)^{2} \left[1-\frac{H}{\Lambda}\,\sin\left(\frac{2\Lambda}{H}\right)\right]=\left(\frac{H}{\dot{\phi}}\right)^2\left(\frac{H}{2\, \pi}\right)^{2}  \left[1-\frac{H}{3\, M_p^2}\,\sin\left(\frac{6\, M_p^2}{H}\right)\right]. \hspace*{10mm}
\label{ps}
\end{eqnarray} 
Fig. \ref{pt_ps} shows the dependence of tonsorial (top panel) and scalar (bottom) density fluctuations, Eqs. (\ref{pt}) and (\ref{ps}), on the scalar field of inflation $\phi$. We show the fluctuations corresponding to the inflation potential and find that the tonsorial and scalar density fluctuations decrease as scalar field of inflation $\phi$ increases. The tonsorial density fluctuations (top panel) corresponding the inflation potentials, $V_1(\phi)$ from Eq. (\ref{eq:mssm}), $V_2(\phi)$ from Eq. (\ref{sdual}) and $V_3(\phi)$ from (\ref{poweri}), look similar. There is a rapid rise at small and a decrease at large $\phi/M_p$. For the inflation potential given in Eq. (\ref{poweri}), $\phi/M_p$ at which the peak takes place is smaller than that for Eq. (\ref{eq:mssm}).  


No systematic comparison can be done for the scalar (bottom panel) density fluctuations of the different inflation potentials. The left-hand and the middle panels shows that the potential, Eqs. (\ref{eq:mssm}) and  (\ref{sdual}), very rapidly decreases with $\phi/M_p$. Then, increasing $\phi/M_p$ does not change the fluctuations. The right-hand panel, Eq. (\ref{poweri}), presents another type of scalar density fluctuations, which remain almost unchanged for a wide range of $\phi/M_p$. Then, the fluctuations are almost damped, at large $\phi/M_p$.

The results corresponding to $\alpha=10^{-2}~$GeV$^{-1}$ are depicted. Exactly the same curves are also obtained at $\alpha=10^{-19}~$GeV$^{-1}$ (not shown here). The earlier value is related to $\alpha_0=10^{17}\, c$ while the latter to $\alpha_0=1\, c$. In light of this, the bounds on $\alpha_0$ seem not affecting the evolution of both tonsorial and scalar density fluctuations with the scalar field of inflation $\phi$. 

Therefore, we can now study the ratio of tensor-to-scalar fluctuations, $r$, which obviously reads \cite{Linde:1982,Liddle:1993,Liddle:1995}
\bea
r &=& \frac{p_t}{ p_s }\,=\,\left(\frac{\dot{\phi}}{H}\right)^2, \label{eq:r}
\eea
relating potential evolution with the Hubble parameter $H$. Corresponding to the tensor-to-scalar fluctuations, a spectral index $n_s$ can be defined
\bea
n_s &=& 1 - \sqrt{\frac{r}{3}}. \label{eq:ns}
\eea
The number of $e$-folds is given by numbers of the Hubble $\mathcal{N}_e \approx 60$ \cite{Liddle:1993} or the integral of the expansion rate,
\bea
\mathcal{N}_e &=& \int _{t_{i}}^{t_{f}} H(t)\, dt = -3 \int _{\phi}^{\phi_{f}}  \frac{H^2}{\partial _{\phi}\, V(\phi)} d\phi ,
\eea
where
\bea
H(t)\, dt &=& 
\frac{H}{\dot{\phi}} d \phi = - 3\, \frac{H^2}{\partial_{\phi}\, V(\phi)} d \phi.
\eea

\begin{figure}[htb]
\includegraphics[width=5.5cm,angle=-90]{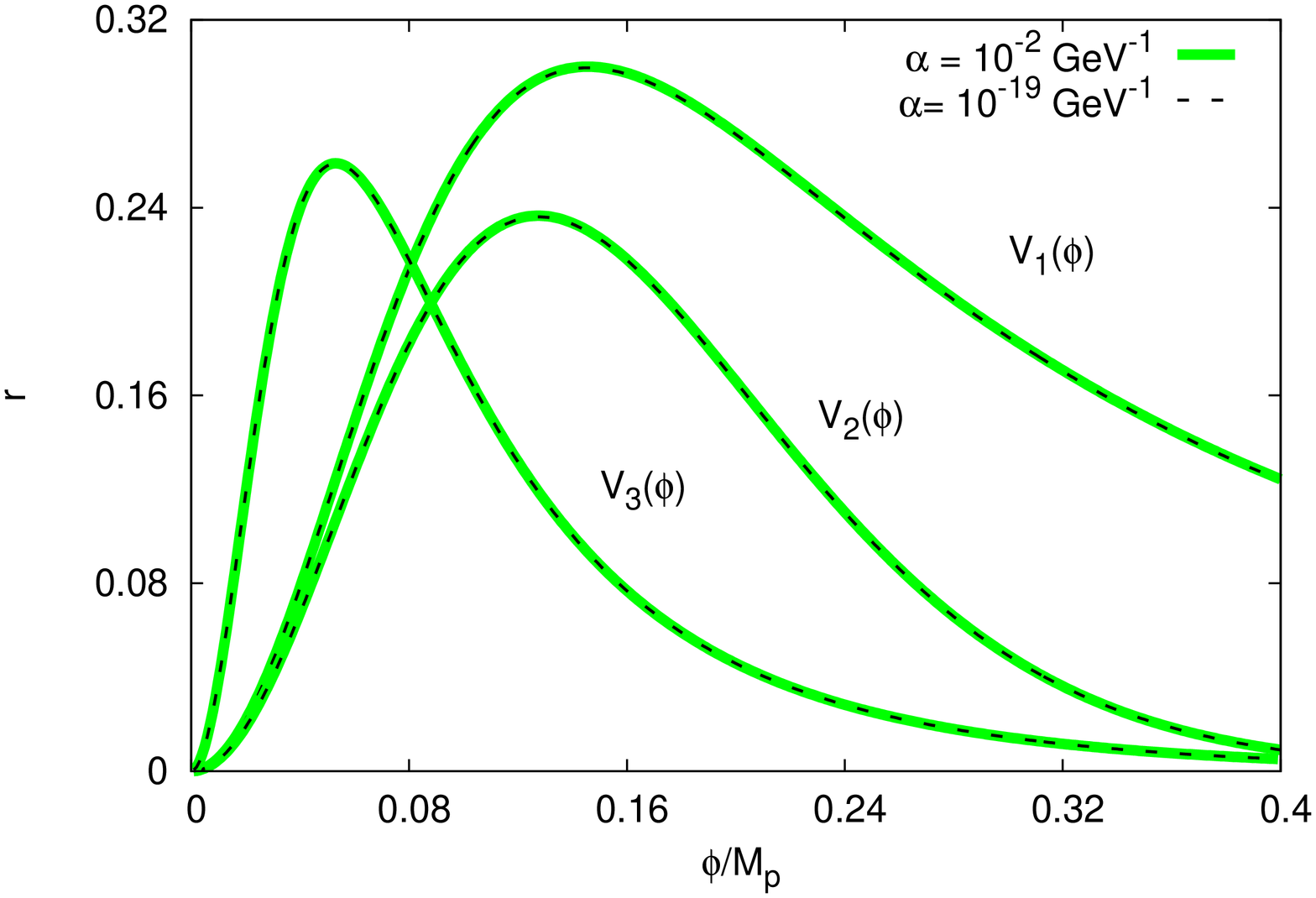}
\includegraphics[width=5.5cm,angle=-90]{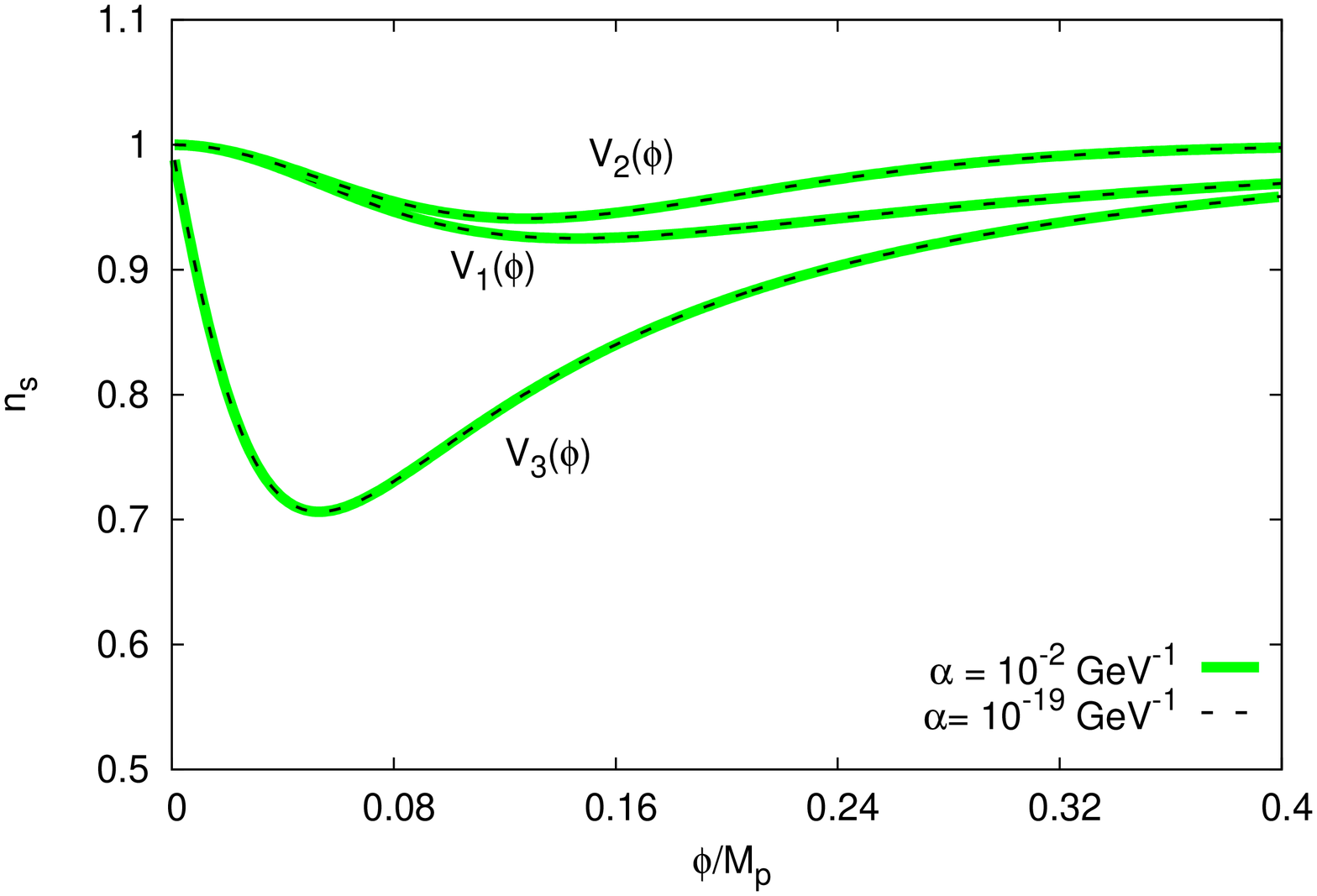}
\caption{(Color online) Left-hand panel shows the ratio of tonsorial-to-scalar density fluctuations, $r$, in dependence on $\phi/M_p$ calculated for the inflation potentials $V_1(\phi)$, $V_2(\phi)$ and $V_3(\phi)$. The right-hand panel gives the spectral index $n_s$ vs. $\phi/M_p$. The dashed curves are evaluated at  $\alpha=10^{-2}~$GeV$^{-1}$, while the solid curves at $\alpha=10^{-19}~$GeV$^{-1}$. 
\label{ratio/spectrial}  
}
\end{figure}

\begin{figure}[htb]
\includegraphics[width=12cm,angle=-90]{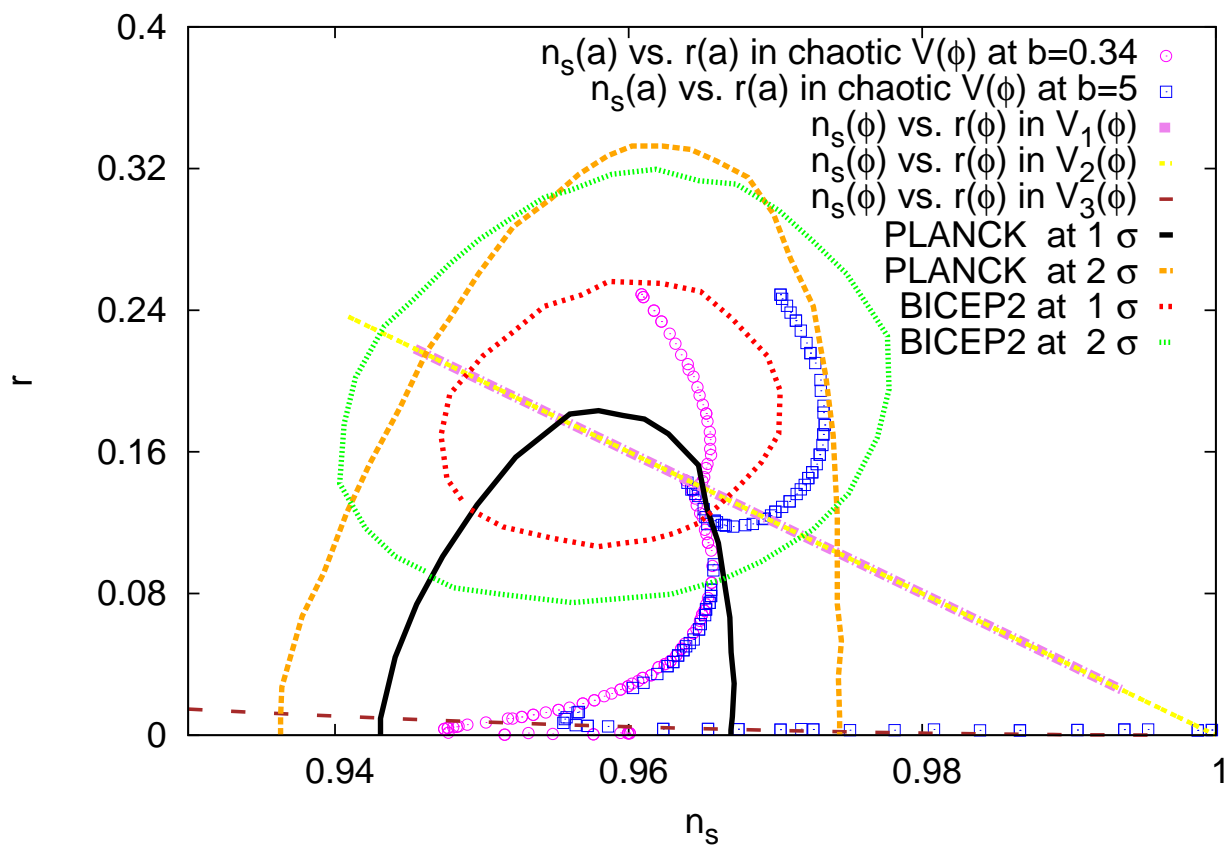}
\caption{(Color online) Contours showing PLANCK and BICEP2 results at $1 \sigma$ and $2 \sigma$ confidence compared with the parametric calculations for $r$ as function of scalar spectral index $n_s$. The parametric calculations for chaotic inflation potential given in Ref. \cite{Andrei Linde:2014}, the square $(b = 0.34)$ and circle $(b =5)$ balls corresponds to $(0.001<\,a\,<0.13)$ and inflation field $\phi \sim 8.2$ are also compared with.
\label{ratio/spectrial2}  
}
\end{figure}

In Fig. \ref{ratio/spectrial}, the left-hand panel shows the ratio of tonsorial to scalar density fluctuations $r$ in dependence on $\phi/M_P$. The dashed curves are evaluated at  $\alpha=10^{-2}~$GeV$^{-1}$, while the solid thick curves at $\alpha=10^{-19}~$GeV$^{-1}$. The earlier value is corresponding to $\alpha_0=10^{17}$ while the latter to $\alpha_0=1$. It is obvious that the bounds on $\alpha_0$ do no affect the ratio of tonsorial to scalar density fluctuations $r$ in dependence on $\phi/M_P$. The behavior of the tonsorial to scalar ratio is limited by the modified Friedmann equation (in the presence of GUP), where the GUP physics is related to the gravitational effect on such model at the Planck scale. The GUP parameter $\alpha$ - appearing in the modified Friedmann equation - should play an important role in bringing the value of $r$ very near to both PLANCK and BICEP2, $r=0.2^{+0.07}_{-0.05}$.  According to Eq. (\ref{modified HH}), $\alpha$ breaks (slows) down the expansion rate, $H$, compared with Fig. \ref{Hubble&scale}. It is obvious that the parameters related to the Gaussian sections of the three curves match nearly perfectly with the results estimated by the PLANCK and BICEP2 collaborations (compare with Fig. \ref{ratio/spectrial2}). 

The right-hand panel of  Fig. \ref{ratio/spectrial} shows the variation of the spectral index, $n_s$, with scalar field for the three inflation potentials, Eqs. (\ref{eq:mssm}), (\ref{sdual}) and (\ref{poweri}). Again, the dashed curves are evaluated at  $\alpha=10^{-2}~$GeV$^{-1}$, while the solid thick curves at $\alpha=10^{-19}~$GeV$^{-1}$. It is obvious that the bounds on $\alpha_0$ do no affect the dependence of spectral index, $n_s$ on $\phi/M_P$. 

Figure \ref{ratio/spectrial2} summarizes the observations of PLANCK \cite{Planck,Planck2} and BICEP2 \cite{BICEP2} collaborations together with the parametric dependence of  spectral index $n_s$ and the ratio $r$. Both parametric quantities are functions of $\phi$, Eqs. (\ref{eq:r}) and (\ref{eq:ns}). {\bf We find that the region of PLANCK at $1\, \sigma$ \cite{Planck,Planck2} and BICEP2 at $1\, \sigma$ \cite{BICEP2} observations for $r$ and $n_s$ is crossed by our parametric calculations for $r$ vs. $n_s$ for two different potentials Eqs. (\ref{eq:mssm}) and (\ref{sdual})}. For the inflation potential, Eq. (\ref{poweri}), the parametric calculations for $r$ vs. $n_s$ are very small. This can be interpreted due to the large minimum in the right-hand panel of Fig. \ref{ratio/spectrial}, which means that main part of $n_s$ calculated for this potential is entirely excluded (out of the range). The related part is obviously very small. The authors of Ref. \cite{Andrei Linde:2014} predict variations of the fluctuations tensor with the spectral index at $55$ {\it e}-folding corresponding to $a$ for the chaotic inflation potential,
\bea
V(\phi)=\frac{m^2 \phi ^2}{2} (1- a\, \phi +a^2 b \phi ^2)^2.
\eea
Our results fit well with the curves of Ref. \cite{Andrei Linde:2014} (open symbols in Fig. \ref{ratio/spectrial2}) which have an excellent agreement with PLANCK and BICEP2 observations. It is worthwhile to highlight that they are deduced using other methods than ours. The main difference is the varying chaotic potential parameters at a constant inflation field. Furthermore,  Ref. \cite{Andrei Linde:2014} gives $n_s (a)$ and $r(a)$, while we are varying various potential with the scalar field at constant potential parameters and estimate $n_s (\phi)$ and $r(\phi)$. The parametric dependence of $n_s (a)$ and $r(a)$ is given in Fig. \ref{ratio/spectrial2}.

{\bf It is apparent that the graphical comparison in Fig. \ref{ratio/spectrial2} presents an excellent agreement between the observations of PLANCK \cite{Planck,Planck2} and BICEP2 \cite{BICEP2} and the parametric calculations, especially for the inflation potentials, Eqs. (\ref{eq:mssm}) and (\ref{sdual}). The agreement is apparently limited to the values given by the parametric calculations, while the observations are much wider. }

Tab. \ref{tab:1} summarizes the results of $r$ and $n_s$ at various scalar fields $\phi/M_p$ for the three inflation potentials, $V_1(\phi)$ from Eq. (\ref{eq:mssm}), $V_2(\phi)$ from Eq. (\ref{sdual}) and $V_3(\phi)$ from (\ref{poweri}). The BICEP2-relevant results are $r$ ranging from $0.15$ from $0.27$ and simultaneously $n_s$ between $0.94$ and $0.98$. It is apparent that the results from $V_3(\phi)$ do not appear in this $r-n_s$ window. The results from $V_1(\phi)$ and $V_2(\phi)$ obviously do. While $V_1(\phi)$ allows a wide range of $\phi$, $V_2(\phi)$ is only relevant for a narrower one. Fig. \ref{ratio/spectrial2} represents this comparison, graphically. 

\begin{table}
\begin{tabular}{||c|c|c||c|c|c||c|c|c||}
\hline \hline 
$\frac{\Delta \phi}{M_p}$ &$ r \,\,\textit{for}\,\, V_{1}({\phi})$ & $n_s \,\,\textit{for}\,\, V_{1}({\phi})$ & $\frac{\Delta \phi}{M_p}$ & $r \,\,\textit{for}\,\, V_{2}({\phi})$ & $n_s \,\,\textit{for}\,\, V_{2}({\phi}$) & $\frac{\Delta \phi}{M_p}$ & $r \,\,\textit{for}\,\, V_{3}({\phi})$ & $n_s \,\,\textit{for}\,\, V_{3}({\phi}$ ) \\ 
\hline \hline 
0.21     &  0.263    & 0.934  & 0.07  &    0.157  &  0.961
 & 0.02 & 0.118 & 0.801 \\ 
\hline 
0.22      &  0.254  &   0.936 & 0.08 &     0.182 &   0.954
&0.03  &    0.192   &   0.746 \\ 
\hline 
0.23       & 0.245   &  0.939 & 0.09 &    0.204   &  0.950
& 0.04  &    0.240   &   0.717 \\ 
\hline 
0.24    &   0.236  &  0.941 & 0.10   &   0.219  &   0.945
& 0.05  &   0.258  &    0.707 \\ 
\hline 
0.25    &   0.227  &  0.943 & 0.11    &  0.230  &   0.942
& 0.06   &  0.255 &     0.708 \\ 
\hline 
0.26   &  0.218 &   0.945 & 0.15 &     0.227   &  0.943
&0.07   &   0.239 &     0.717 \\ 
\hline 
0.27    &    0.210  &  0.947 & 0.16    &  0.218   &  0.945
 & 0.08   &   0.218  &    0.730 \\ 
\hline 
0.29    &   0.193  &  0.951 & 0.17  &    0.207   &  0.948
 & 0.09  &    0.194    &  0.745 \\ 
\hline 
0.3     &   0.185  &  0.953 & -- & -- &-- &0.1   &   0.172    &  0.761 \\ 
\hline 
0.32    &    0.171 &   0.957 & -- & -- & -- & 0.11   &   0.151    &  0.775 \\ 
\hline 
0.33    &   0.164 &  0.959 & -- & -- & -- & -- &--& -- \\ 
\hline 
0.35    &   0.151   & 0.962 &-- & -- & -- & -- &--& -- \\ 
\hline 
0.36  &   0.145   & 0.963 & -- & -- & -- & -- & -- & -- \\ 
\hline\hline 
\end{tabular}
\caption{The ratio of tonsorial to scalar density, the fluctuations $r$  and the spectral index $n_s$ associated with the scalar field for different inflation potentials, $V_1(\phi)$ from Eq. (\ref{eq:mssm}), $V_2(\phi)$ from Eq. (\ref{sdual}) and $V_3(\phi)$ from (\ref{poweri}). \label{tab:1}}
\end{table}
 

\section{Discussion and conclusions} 
\label{summary}

The BICEP2 results announced on March 17, 2014 made the physicists around the globe having another view about the evidence of Universe and its expansion, especially at about the inflation era. The cosmic inflation is based on the assumption that an extreme inflationary phase should take place after the Big Bang (at about the Planck time). Thus, the Universe should expand at a superluminal speed. On the other hand, the inflation would result from a hypothetical field acting as a cosmological constant to produce an acceleration expansion of the Universe. 

Argumentation about the applicability of GUP on the inflation era will be elaborated in Append. \ref{app:appl}. Due to the very high energy (quantum or Planck scale), the Heisenberg uncertainty principle should be modified in terms of momentum uncertainty.  The QG approach in form of GUP appears in the modified Friedmann equation - in terms of $\alpha$. This term reduces the Hubble parameter, which appears in the denominator of the ratio of tonsorial-to-scalar density fluctuations. Thus, the fluctuations ratio increases due to decreasing $H$. The  fluctuations ratio $r$ has been evaluated as function of the spectral index $n_s$. We found that the calculations match well with the PLANCK and BICEP2 observations. This is the main conclusion of the present work. We believe that the results point to the importance of quantum correlation during the inflation era. 

The estimation of the $n_s (a)$ and $r(a)$ at $55$ {\it e}-folds for a chaotic potential for different values of $b$ and varying inflation as function of  $a$. The parameters $b = 0.34$ (open squares) and $b =5$ (open circles) are corresponding to $(0.001<a<0.13)$ and inflation field $\phi \sim 8.2ss$ \cite{Andrei Linde:2014}. The authors predict the variation of the fluctuation tensor with the spectral index. The best curves in Ref. \cite{Andrei Linde:2014} agree well with PLANCK and BICEP2.  These are deduced using another method, varying $a$ and selecting out the suitable scalar field. The main difference with our method is the is the varying chaotic potential parameters at constant inflation field. We vary the inflation potential with the scalar field at constant  potential parameters.   

We have reviewed different inflationary potentials and estimated the modifications of the Friedmann equation due to the GUP approach. We found that 
\begin{itemize}
\item the first potential, Eq. (\ref{eq:mssm}), gives a  power law of the scalar  inflation-field. This is based on certain minimal supersymmetric extensions of the standard model \cite{allahverdi-2006}. 
\item The second potential, Eq. (\ref{sdual}), hypothesizes that the potential should be invariant under the $\mathcal{S}$-duality constraint $g\rightarrow 1/g$, or $\phi \rightarrow -\phi$, where $\phi$ is the dilation/inflation and $g \approx \exp \left(\phi/M\right)$ \cite{sdual}. The $\mathcal{S}$-duality had its roots in the Dirac quantization condition for the electromagnetic field. Thus, it should be equivalence to the description of quantum electrodynamics as either a weakly coupled theory of electric charges or a strongly coupled theory of magnetic monopoles \cite{JSchwarz:2002}. The latter, Eq. (\ref{poweri}) appeared in an exponential form with a power-law inflation field. These inflationary potentials seem to agree well with of the observations of PLANCK and BICEP2 collaborations at  different $1\,\sigma$ and $2\, \sigma$. In the range of spectral index and fluctuation ratio. 
\item The potential, Eq. (\ref{poweri}) disagrees. Few remarks are now in order. The agreement should be limited to the values given by the parametric calculations. The PLANCK and BICEP2 observations are much wider but have uncertainties in $r$ of order $25\%$. We have presented through a conceivable way the effects of reasonably-sized GUP parameter of our estimation for $r$.
\end{itemize}
We conclude that depending on the inflation potential $V(\phi)$ and the scalar field, $\phi$, the GUP approach seems to reproduce the BICEP2 observations $r=0.2 _{-0.05}^{+0.07}$, which also have been fitted by using $55$ {\it e}-folds for a chaotic potential for varying inflation and seem to agree well with the upper bound value corresponding to PLANCK and to WMAP9 experiment.

\appendix

\section{Generalized uncertainty principle (GUP)}
\label{GUPy}

\subsection{Minimal length uncertainty and maximum measurable momentum}
\label{GUPH}

The commutator relation \cite{advplb,Das:2010zf,afa2},  which are consistent with the string theory, the black holes physics and DSR leads to
 \bea
\left[\hat{x_{i}}, \hat{p_{j}} \right]=i \hbar \left[ \delta_{i j} -\alpha \left( p \delta _{i j} +\frac{p_{i} p_{j}}{p} \right)+\alpha ^{2} \left(p^{2} \delta_{ij} +3 p_{i} p_{j} \right)\right], \label{ali1}
\eea
implying a minimal length uncertainty and a maximum measurable momentum when implementing convenient representation of the commutation relations of the momentum space wave-functions \cite{amir,Tawfik:BH2013}.  The constant coefficient $\alpha=\alpha_0 /(M_p\, c)=\alpha_0\, l_p/\hbar$ is  referring to the quantum-gravitational effects on the Heisenberg uncertainty principle. The momentum $\hat{p}_{j}$ and the position $\hat{x}_{i}$ operators are given as  
\bea 
\hat{x}_{i}\, \Psi (p) &=& x_{0 i}(1 -\alpha \,p_{0} +2\, \alpha^{2}\, p_{0}^{2})\, \Psi (p), \label{amm1}\nn \\
\hat{p}_{j}\, \Psi (p) &=&  p_{0 j}\, \Psi (p). \label{amm2}
\eea
We notice that $p_{0}^{2}=\sum_{j}^3p_{0 j}\,p_{0 j}$ satisfies the canonical commutation relations $\left[x_{0 i},\, p_{0 j}\right]=i\,\hbar\,\delta_{i j}$. Then, the minimal length uncertainty \cite{advplb,Das:2010zf,afa2}  and maximum measurable momentum \cite{amir,Tawfik:BH2013}, respectively, read
\bea 
\Delta x &\geq & (\Delta x)_{min} \approx \hbar \alpha, \nn \\
p_{max} &\approx & \frac{1}{4 \alpha},
\eea
where the maximum measurable momentum agrees with the value which was obtained in the doubly special relativity (DSR) theory \cite{DSR,Tawfik:BH2013}. By using natural units, the one-dimensional uncertainty reads \cite{advplb,Das:2010zf,afa2}
\bea 
\Delta x\, \Delta p &\geq & \frac{\hbar}{2} \left(1-2\, \alpha\, \Delta p +4\, \alpha^{2} \Delta p^{2}  \right). \label{ali3}
\eea
This representation of the operators product satisfies the non-commutative geometry of the spacetime \cite{amir}
\bea
\left[\hat{p_{i}},\hat{p_{j}}\right]&=&0,\nn \\
\left[\hat{x_{i}},\, \hat{x_{j}} \right] &=& - i\, \hbar\, \alpha\, \left(4\, \alpha - \frac{1}{P}\right)\, \left(1 - \alpha\, p_{0} +2 \alpha^{2}\, \vec{p_{0}}^{2}\right)\; { \hat L_{i j}}. 
\eea
The rotational symmetry does not break by the commutation relations \cite{amir,Kempf95}. In fact, the rotation generators can still be expressed in terms of position and momentum operators as \cite{amir,Tawfik:BH2013}
\bea
L_{i j} &=& \frac{\hat X_{i}\,  \hat P_{j} -\hat{X}_{j}\, \hat{P}_{i}}{1 -\alpha\, p_{0} +2\, \alpha^{2}\, \vec{p_{0}}^{2}}.
\eea

\subsection{Applicability of GUP to the cosmic inflation} 
\label{app:appl}

The quantum aspects of the gravitational fields can emerge in the limit, where strong, weak and electromagnetic interactions can be distinguished from each other. In the view of {\it gedanken} experiments that have been designed to measure the apparent horizon area of a black hole in QG \cite{Maggiore94}, the uncertainty relation is found preformed \cite{Maggiore93}. The deformed or modified Heisenberg algebra, which was suggested to investigate GUP, introduces a relation between QG and Poincare algebra \cite{Maggiore94}. Nevertheless, GUP given in quadratic forms \cite{Maggiore93,Kempf}, which fits well with the string theory and the black hole physics and introduces a minimal length uncertainty and additional linear terms of momenta \cite{DSR} agrees also well with DSR and assumes that the momenta approach maximum value at very high energy (Planck scale) \cite{DSR}.

There are several observations supporting the concept of GUP approaches and offering a possibility of studying the influence of the minimal length on the properties of a wide range of physical systems, especially at quantum scale  \cite{Maggiore93,Maggiore94,Scardigli}. The effects of linear GUP approach have been studied on compact stars \cite{Ali:2013ii}, Newtonian law of gravity \cite{Ali:2013ma}, inflationary parameters and thermodynamics of the early Universe \cite{Tawfik:2012he}, Lorentz invariance violation \cite{Tawfik:2012hz} and measurable maximum energy and minimum time interval \cite{DahabTaw}. Furthermore, the effects of QG on the quark-gluon plasma (QGP) are studied  \cite{Elmashad:2012mq}. It was found that the GUP can potentially explain the small observed violations of the weak equivalence principle in neutron interferometry experiments \cite{exp}. Also, it was suggested \cite{nature2012} that GUP can be measured directly in Quantum Optics Lab  \cite{Das:2010zf,afa2}. The current researches of the quantum problems in the presence of gravitational field at very high energy near to the Planck scale implies new physical laws and even corrections of the spacetime of our Universe \cite{Kempf}. The quantum field theory in curved background can be normalized by introducing a minimal observable length as an effective cut-off in ultraviolet domain \cite{Kempf}. It is conjectured that the string can't probe distances smaller than its own length. For cosmic inflation, is the expansion of space in the early Universe. The inflationary epoch lasted from $10^{-36}$ seconds after the Big Bang to sometime between $10^{-33}$ and $10^{-32}$ seconds near to Planck scale. Following the inflationary period, the universe continues to expand, but at a less accelerated rate. Actually, the GUP at very high energy Planck scale would likely be applicable to contract these approach interpret of the quantum study of the inflationary of the universe. 

\section{Modified Dispersion Relation (MDR)}
\label{app:mdr}

Various observations support the conjectured that the Lorentz invariance might be violated. The velocity of light should differ from $c$. Any tiny adjustment leads to  modification of the energy-momentum relation and modifies the dispersion relation in vacuum state by $\delta \textit{v}$ \cite{Glashow:1,Glashow:2,Glashow:3,Amelino98}. In particular, at the Planck scale, the modifications of energy-momentum dispersion relation have been considered in Refs. \cite{lqgDispRel1,lqgDispRel2}. Two functions $p(E)$ as expansions with leading Planck-scale correction of order $L_p\, E^3$ and $L_p^2 E^4$ respectively, reads \cite{Amelino2004b},
\bea
\vec{p}^2 &\simeq & E^2 - m^2 + \alpha_1\, L_p\, E^3, \label{disprelONE} \\
\vec{p}^2 &\simeq & E^2 - m^2 + \alpha_2\, L_p^2\, E^4.
\label{disprelTWO}
\eea
These are valid for a particle of mass $M$ at rest, whose position is being measured by a procedure involving a collision with a photon of energy $E$ and momentum $p$. Since the relations are originated from Heisenberg uncertainty principle for position with precision $\delta x$, one should use a photon with momentum uncertainty $\delta p \ge 1/\delta x$. Based on the argument of Ref. ~\cite{landau} in loop QG, we convert $\delta p \ge 1/\delta x$ into $\delta E \ge 1/\delta x$. By using the special-relativistic dispersion relation and $\delta E \ge 1/\delta x$, then $M \ge \delta E$. If indeed loop QG hosts a Planck-scale-modified
dispersion relation, Eq. (\ref{disprelTWO}), then$\delta p_\gamma \ge 1/\delta x$ and this required that \cite{Amelino2004b},
\begin{equation}
M  \ge \frac{1}{\delta x} \left(1 - \alpha_2 \frac{3 L_p^2}{2 (\delta x)^2}\right). \label{dis}
\end{equation}
These results apply only to the measurement of the position of a particle at rest \cite{landau}.  We can generalize these results to measurement of the position of a particle of energy $E$.
\begin{itemize}
\item In case of standard dispersion relation,  one obtains that $E  \ge 1/\delta x$ as required for a  linear dependence of entropy on area, Eq. (\ref{disprelTWO})
\item For the dispersion relation, Eq. (\ref{disprelTWO})  
\bea
E  \ge \frac{1}{\delta x} \left(1- \alpha_2 \frac{3 L_p^2}{2 (\delta x)^2}\right).
\eea
The requirements of these derivation lead in order of correction of log-area form.
\item Furthermore, 
\bea
E  \ge \frac{1}{\delta x} \left( 1+ \alpha_1 \frac{L_p}{\delta x}\right).
\eea
\end{itemize}
In case of string theory, the {\it ''reversed Bekenstein argument''} leads to  quadratic GUP, that fits well with the string theory \cite{venegross} and black holes physics,
\begin{equation}
\delta x  \ge \frac{1}{\delta p} + \lambda_s^2  \delta p. \label{gup}
\end{equation}
The scale $\lambda_s$ in Eq. (\ref{gup}) is an effective string length giving the characteristic length scale which be identical with Planck length. Many researches of loop QG \cite{lqgDispRel1,lqgDispRel2} support the possibility of the existence of a minimal length uncertainty and a modification in the energy-momentum dispersion relation at Planck scale. 


\end{document}